\documentclass[pra,aps,superbib,superscriptaddress,twocolumn]{revtex4-1}

\usepackage{amsmath,bm}
\usepackage{amstext}
\usepackage{epsfig}
\usepackage{xcolor}
\usepackage{subfig}
\usepackage{graphicx}
\usepackage{multirow}
\usepackage{array}
\usepackage{tikz,pgfplots}
\usepackage{MnSymbol}
\definecolor{dgreen}{rgb}{0,.5,0}
\definecolor{dred}{rgb}{.7,.0,.0}
\def\ddroit{{d}}
\newcommand{\be}{\begin{eqnarray}}
\newcommand{\ee}{\end{eqnarray}}

\begin{document}

\title{
Unified formulation of fundamental and optical gap problems in 
density-functional theory for ensembles
}

\author{Bruno Senjean}
\thanks{Corresponding author}
\email{senjean@unistra.fr}
\affiliation{Laboratoire de Chimie Quantique,
Institut de Chimie, CNRS/Universit\'e de Strasbourg,
4 rue Blaise Pascal, 67000 Strasbourg, France}

\author{Emmanuel Fromager}
\affiliation{Laboratoire de Chimie Quantique,
Institut de Chimie, CNRS/Universit\'e de Strasbourg,
4 rue Blaise Pascal, 67000 Strasbourg, France}

%%%%% Abstract %%%%%

\begin{abstract}

Solving the fundamental and
optical gap problems, which yield information about charged and neutral excitations in
electronic systems, is one of the biggest challenge in
density-functional theory (DFT). Despite their intrinsic  
difference, we show that the two problems can be made formally identical
by introducing a universal and canonical ensemble
weight dependent exchange-correlation (xc) density functional. The weight dependence of the xc energy turns out to be the key ingredient for
describing the infamous derivative discontinuity and represents a new path for
its approximation.
\end{abstract}

\maketitle

\section{Introduction}
Kohn--Sham (KS) density-functional theory (DFT)~\cite{hktheo,KS} has
become over the last two decades the method
of choice for electronic structure calculations in molecules and solids.
This great success relies on the mapping of the 
physical interacting problem onto
a non-interacting one, thus leading to a dramatic reduction of the computational cost
in contrast to more involved many-body approaches. In DFT, the
exchange and correlation (xc) contributions to the two-electron repulsion
energy are universal functionals of the electron density. Among the numerous
properties of interest
are the fundamental and optical gaps which describe charged and neutral electronic excitations, respectively.
The accurate description of these quantities 
is crucial in the design of new
nanodevices such as molecular junctions, for example.\\

A nice feature of DFT is that these gaps and, more generally, any
excitation energy, can be related to the KS
orbital energies. 
Nevertheless, making this relation as explicit as
possible remains a true challenge. Indeed, in the standard formulation of
DFT, it is crucial to describe correctly not only the KS orbital energies but
also
the discontinuous behavior of the xc potential (i.e. the derivative of the xc
energy with respect to the density) induced by the 
excitation process, whether it is neutral~\cite{PRA_Levy_XE-N-N-1,yang2014exact} or not~\cite{perdew1983physical,
DFT_energygap,
parr1983some,
godby1987trends,
godby1988self,
nesbet1997fractional,
harbola1998differential,
chan1999fresh,
teale2008orbital,
yang2012derivative,
gould2014kohn,
dimitrov2016exact,
benitez2016kohn,
hodgson2017derivative}. Unfortunately, standard xc functionals do not
exhibit such a derivative discontinuity (DD) which explains why, in
practice, both chemistry and physics communities have turned
to generally more expensive ``post-DFT''
methods like time-dependent (TD) DFT~\cite{PRL84_RGth} for the computation of neutral
excitations and, for the charged ones, 
to DFT+$U$~\cite{LDA+U_reformulation_localorbital,
DFT_Transition-metal_selfHubbardUapproach,
DFT+U_transitionmetals,
anisimov1991band,
liechtenstein1995density,cococcioni2005linear}, hybrid functionals~\cite{imamura2011linearity,                     
atalla2016enforcing,stein2010fundamental,
stein2012curvature} or the even more involved Green's function-based
methods like GW~\cite{bruneval2012ionization,                                              
bruneval2012benchmarking,                                                            
jiang2015first,
pacchioni2015first,
ou2016comparison,
reining2018gw}.\\

The Gross--Oliveira--Kohn (GOK) DFT for canonical ensembles~\cite{PRA_GOK_RRprinc,
PRA_GOK_EKSDFT,
GOK3} has
gained increasing interest in recent years as it provides a rigorous way
to extract neutral excitation energies in a completely time-independent
framework~\cite{yang2014exact,yang2017direct,gould2017hartree,gould2018charge,deur2017exact,deur2018exploring}. In this context, the DD is automatically described through
the ensemble weight dependence of the xc
functional~\cite{PRA_GOK_EKSDFT,PRA_Levy_XE-N-N-1,yang2014exact}, which is extremely
appealing. The method is in principle much cheaper computationally than standard
approximate TD-DFT
and, in contrast to the latter, it allows for the description of
multiple electron excitations~\cite{yang2017direct,gould2017hartree}.\\

Turning to charged 
excitations, we know from the seminal work of Perdew and Levy~\cite{perdew1983physical} that it is
in principle sufficient to extend the domain of definition of the conventional
xc functional to fractional electron numbers in order to account for the
DD. In practice, the task is far from trivial and, despite significant progress~\cite{
cohen2008insights,
cohen2008fractional,
mori2008localization,
sai2011hole,
kronik2012excitation,
refaely2012quasiparticle,
miranda2016fractional,
li2017extending,
perdew2017understanding,zhou2017unified,
zheng2011improving,
zheng2013nonempirical,
li2015local,
andrade2011prediction,gorling2015exchange,
thierbach2017accurate}, no clear strategy has emerged over the past
decades. Quite recently, Kraisler and
Kronik made the formal connection between non-neutral excitations
and GOK-DFT more explicit by introducing a grand canonical
ensemble weight, thus paving the way to the construction of more
reliable xc functionals for ionization and affinity processes~\cite{kraisler2013piecewise,kraisler2014fundamental}.
Unfortunately, as the total (fractional) number of electrons varies with
the weight, the analogy with GOK-DFT can only be partial.\\

The purpose of this work is to prove that, with an appropriate
choice
of grand canonical ensemble, informations about non-neutral excitations
can be extracted, in principle exactly, from a
canonical
(time-independent) formalism.
As a
remarkable result, the optical and
fundamental gap problems become
formally identical, even though the physics they
describe is completely different. Although it had not been
realized yet, advances in
GOK-DFT should therefore be beneficial to the description of fundamental
gaps too. 
The paper is organized as follows. An in-principle-exact
single-weight ensemble DFT is derived for the fundamental gap in
Sec.~\ref{subsec:one-weight_edft},  
in analogy with GOK-DFT. A two-weight generalization is then
introduced in Sec.~\ref{2-weight_edft}, in order to extract both
ionization potential and electron affinity separately. The theory, which
is referred to as $N$-centered ensemble DFT, is then applied in
Sec.~\ref{sec:Hubbard_dimer} to the
simple but nontrivial asymmetric Hubbard dimer, as a proof of concept.
Conclusions and perspectives regarding, in particular, the construction
of {\it ab initio} weight-dependent density-functional approximations
are given in Sec.~\ref{sec:conclusions}.\\  

\section{Theory}

\subsection{Single-weight $N$-centered ensemble
DFT}\label{subsec:one-weight_edft}

In the conventional DFT formulation of the fundamental gap
problem, a grand canonical ensemble consisting of $(N-1)$- and
$N$-electron ground states 
is considered,
thus leading to a total number of electrons that can be fractional. By
analogy with the time-ordered one-particle Green's function, which
contains information about the $(N-1)$-, $N$-, and $(N+1)$-electron
systems, we propose instead to  
consider what we will refer to as an {\it $N$-centered} grand canonical
ensemble. The latter will be characterized by a {\it central} number $N$ of
electrons and an ensemble weight $\xi$, in the range $0\leq \xi\leq
1/2$, that is assigned to {\it both} $(N-1)$- and $(N+1)$-electron
states. In the following, the ensemble will be denoted as $\{ N, \xi\}$.   
It is formally described by the following ensemble density matrix operator, 
\begin{eqnarray}\label{eq:n-centered-dens_mat_op}
\hat{\Gamma}^{\{N,\xi\}}=\xi\hat{\Gamma}^{N_-}
+\xi\hat{\Gamma}^{N_+}+(1-2\xi)\hat{\Gamma}^{N},
\end{eqnarray}
which is a convex combination of $\mathcal{N}$-electron density matrix
operators $\hat{\Gamma}^\mathcal{N}$ with $\mathcal{N}\in\left\{N_-,N,N_+\right\}$. Note that, for
sake of compactness, we used the shorthand notations $N_-=N-1$ and
$N_+=N+1$ (not to be confused with left- and right-hand limits). If pure states are used (which is not compulsory)
then $\hat{\Gamma}^{\mathcal{N}}= 
\left\vert\Psi^{\mathcal{N}}\rangle\langle\Psi^{\mathcal{N}}\right\vert$
where $\Psi^{\mathcal{N}}$ is an $\mathcal{N}$-electron many-body
wavefunction. 
Although
the $N$-centered ensemble describes the addition ({\it and} removal) of an
electron to (from)
an $N$-electron system, the corresponding $N$-centered ensemble density,
\be\label{eq:N-centered_dens}
\hspace{-0.4cm}n_{\hat{\Gamma}^{\{N,\xi\}}}(\mathbf{r})&=&
\xi n_{\Psi^{N_-}}(\mathbf{r})
+\xi n_{\Psi^{N_+}}(\mathbf{r})+(1-2\xi)n_{\Psi^{N}}(\mathbf{r})
,
\ee
integrates to the central {\it integral} number of electrons 
$N$. Thus we generate a canonical density
from a grand canonical ensemble. 
This is the
fundamental difference between conventional DFT for open systems and  
the $N$-centered ensemble DFT derived in the following. 
Note that, in a more chemical language, the
deviation of the
$N$-centered ensemble density from the $N$-electron one
$n_{\Psi^{N}}(\mathbf{r})$ is nothing but the difference between right
and left Fukui functions~\cite{parrFukui84} scaled by the ensemble weight $\xi$.
\\

For a given
external potential $v_{\rm ext}(\mathbf{r})$, we can construct, in
analogy with Eq.~(\ref{eq:n-centered-dens_mat_op}), the following $N$-centered ground-state ensemble
energy, 
\begin{eqnarray}\label{eq:ens_ener}
E_0^{\{N,\xi\}}&=&
\xi E_0^{N_-}+\xi E_0^{N_+}+(1-2\xi)E^N_0,
\end{eqnarray}
where $E_0^{\mathcal{N}}$ is the $\mathcal{N}$-electron ground-state
energy of $\hat{H}=\hat{T}+\hat{W}_{\rm
ee}+\int\ddroit\mathbf{r}\;v_{\rm ext}(\mathbf{r})\hat{n}(\mathbf{r})$,
and $\hat{n}(\mathbf{r})$ is the density operator.
The operators $\hat{T}$ and $\hat{W}_{\rm
ee}$ describe the electronic kinetic and repulsion
energies, respectively. 
Note that the $N$-centered ground-state 
ensemble energy is linear in $\xi$ and its slope is nothing but the
fundamental gap. From the following extension of the Rayleigh--Ritz variational
principle,% to $N$-centered ground-state ensembles,
\begin{eqnarray}\label{eq:ens_var_principle}
E_0^{{\{N,\xi\}}}=
 \min_{\hat{\Gamma}^{\{N,\xi\}}} 
{\rm Tr}\left[\hat{\Gamma}^{\{N,\xi\}}\hat{H}\right]
={\rm Tr}\left[\hat{\Gamma}_0^{\{N,\xi\}}\hat{H}\right],
\end{eqnarray}
where Tr
denotes the trace, we conclude that the Hohenberg--Kohn theorem~\cite{hktheo} applies to 
$N$-centered ground-state ensembles
for any {\it fixed} value of $\xi$. Let us
stress that, unlike in DFT for fractional electron numbers, the
one-to-one correspondence between the $N$-centered ensemble density and
the external potential holds {\it up to a
constant}, simply because the former density integrates to a fixed
central number
$N$ of electrons. We can therefore extend DFT to $N$-centered
ground-state ensembles and obtain the energy variationally as follows,  
\begin{eqnarray}\label{eq:ens_var_principle_dens_vext}
E_0^{{\{N,\xi\}}}=\min_{n\rightarrow N}\left\{F^{{\{N,\xi\}}}[n]+
\int\ddroit{\mathbf{r}}\,v_{\rm ext}(\mathbf{r})n(\mathbf{r})\right\},
\end{eqnarray}
where the minimization is restricted to densities that integrate to $N$.
As readily seen from Eq.~(\ref{eq:ens_ener}), conventional ($N$-electron)
ground-state DFT is recovered when $\xi=0$. The analog of the Levy--Lieb
functional for $N$-centered
ground-state ensembles reads
\begin{eqnarray}\label{eq:LL_interacting}
F^{{\{N,\xi\}}}[n]=\min_{\hat{\Gamma}^{\{N,\xi\}}\rightarrow n}
{\rm Tr}\left[\hat{\Gamma}^{\{N,\xi\}}\left(\hat{T}+\hat{W}_{\rm
ee}\right)\right]
,
\end{eqnarray}
where the minimization is restricted to $N$-centered ensembles with
density $n$. Let us consider the KS decomposition, 
\begin{eqnarray}\label{eq:KS_decomp_Fens}
F^{{\{N,\xi\}}}[n]=T_{\rm s}^{{\{N,\xi\}}}[n]+E^{{\{N,\xi\}}}_{\rm Hxc}[n],
\end{eqnarray}
where 
\be\label{eq:Tsxi} 
T_{\rm s}^{{\{N,\xi\}}}[n]=\min_{\hat{\Gamma}^{\{N,\xi\}}\rightarrow n}
{\rm Tr}\left[\hat{\Gamma}^{\{N,\xi\}}\hat{T}
\right]
\ee
 is the non-interacting 
kinetic energy contribution and 
\be\label{Nc-Hxc_equal_H_plus_xc}
E^{{\{N,\xi\}}}_{\rm Hxc}[n]
=
 \frac{1}{2} \iint
\ddroit{\bf r} \ddroit{\bf r'}
~ \dfrac{n({\bf r})n({\bf r'})}{\mid{{\bf r}-{\bf r'}}\mid}
+
E^{{\{N,\xi\}}}_{\rm xc}[n]
\ee
is the $\xi$-dependent analog of the Hartree-xc (Hxc) functional for $N$-centered ground-state
ensembles. 
Note that, even though the electronic excitations described in $N$-centered ensemble DFT and
GOK-DFT~\cite{PRA_GOK_EKSDFT} are completely different, the two theories
are formally identical. Interestingly, as proved in
Appendix~\ref{sec:Ts_proof}, the non-interacting kinetic
energy functionals used in both theories are actually
equal. This is  
simply due to the fact that, in a non-interacting system, the fundamental and
optical gaps boil down to the same quantity. This is of course not the case for interacting
electrons, which means that each theory requires the construction of a
{\it specific} weight-dependent xc functional.\\ 

For that purpose, we
propose to extend to $N$-centered ground-state ensembles the generalized adiabatic
connection formalism for ensembles (GACE) which was originally introduced in the context of
GOK-DFT~\cite{franck2014generalised,
deur2017exact}. In contrast to standard DFT for grand canonical
ensembles~\cite{kraisler2013piecewise}, the ensemble weight
$\xi$ can in principle vary in $N$-centered ensemble DFT while holding 
the density {\it constant}. 
Consequently, we can derive the following
GACE formula,
\begin{eqnarray}\label{eq:GACE}
E^{{\{N,\xi\}}}_{\rm xc}[n]=E_{\rm
xc}[n]+\int^\xi_0\ddroit\alpha\;\Delta^{\{N,\alpha\}}_{\rm xc}[n],
\end{eqnarray}
where, unlike in conventional adiabatic connections~\cite{Nagy_ensAC}, 
we integrate over the ensemble weight rather than
the two-electron interaction strength. The GACE integrand $\Delta^{\{N,\alpha\}}_{\rm xc}[n]
=\partial E^{{\{N,\alpha\}}}_{\rm
xc}[n]/\partial \alpha $ quantifies 
the deviation of the $N$-centered ground-state ensemble xc functional from the
conventional (weight-independent) ground-state one $E_{\rm
xc}[n]=E^{\{N,\xi=0\}}_{\rm xc}[n]$. As shown in Appendix~\ref{appendix:single_weight_GACE_int}, the GACE integrand is simply equal to the difference in
fundamental gap between the interacting and non-interacting systems
with $N$-centered ground-state ensemble density $n$ (and weight
$\alpha$):
\begin{eqnarray}\label{eq:GACE_integrand}
\Delta^{\{N,\alpha\}}_{\rm xc}[n]
=E^{\{N,\alpha\}}_{\rm g}[n]
-
\left(\varepsilon^{\{N,\alpha\}}_{\rm L}[n]
- \varepsilon^{\{N,\alpha\}}_{\rm H}[n]\right).
\end{eqnarray}
\\

Let us now return to the variational ensemble energy
expression in Eq.~(\ref{eq:ens_var_principle_dens_vext}). Combining the
latter with  
Eqs.~(\ref{eq:KS_decomp_Fens}) and (\ref{eq:Tsxi}) leads to 
\begin{eqnarray}\label{eq:min_KS-eDFT}
E_0^{{\{N,\xi\}}}=
 \min_{\hat{\Gamma}^{\{N,\xi\}}}
\Big\{
{\rm
Tr}\left[
\hat{\Gamma}^{\{N,\xi\}}\left(\hat{T}+\hat{V}_{\rm ext}\right)\right]
+
E^{{\{N,\xi\}}}_{\rm Hxc}[n_{\hat{\Gamma}^{\{N,\xi\}}}]
\Big\}
, \nonumber \\
\end{eqnarray}
where $\hat{V}_{\rm ext}=\int\ddroit\mathbf{r}\;v_{\rm
ext}(\mathbf{r})\hat{n}(\mathbf{r})$. Note that the minimizing 
density matrix operator $\hat{\Gamma}_{\rm KS}^{\{N,\xi\}}$
 in
Eq.~(\ref{eq:min_KS-eDFT}) is the non-interacting $N$-centered
ground-state ensemble one whose density equals the physical interacting
one $n_{\hat{\Gamma}_0^{\{N,\xi\}}}(\mathbf{r})$. It can be constructed from a {\it single} set of orbitals which
fulfill the following self-consistent KS equations [the latter are simply
obtained from the stationarity condition associed to Eq.~(\ref{eq:min_KS-eDFT})],
\begin{eqnarray}\label{eq:mono_selfcons}
&&\left[ -\dfrac{\nabla^2}{2} + 
v^{{\{N,\xi\}}}_{\rm
KS}
(\mathbf{r})
\right] \varphi_i^{\{N,\xi\}}(\mathbf{r})
%\nonumber\\
%&&
 = \varepsilon_i^{\{N,\xi\}}\varphi^{\{N,\xi\}}_i(\mathbf{r})
,
\end{eqnarray}
where 
$v^{{\{N,\xi\}}}_{\rm
KS}
(\mathbf{r})=
v_{\rm ext}(\mathbf{r}) +
v^{{\{N,\xi\}}}_{\rm
Hxc}[n_{\hat{\Gamma}_{0}^{\{N,\xi\}}}]
(\mathbf{r})
$
and
$v^{{\{N,\xi\}}}_{\rm Hxc}[n](\mathbf{r})=
{\delta E^{{\{N,\xi\}}}_{\rm
Hxc}[n]}/{\delta
n(\mathbf{r})}
$. In the particular case of pure non-interacting
$\mathcal{N}$-electron states, 
\begin{eqnarray}
n_{\hat{\Gamma}_0^{\{N,\xi\}}}(\mathbf{r})&=&\sum^{N_-}_{i=1}\left\vert\varphi^{\{N,\xi\}}_i(\mathbf{r})\right\vert^2 
+(1-\xi)\left\vert\varphi^{\{N,\xi\}}_{\rm H}(\mathbf{r})\right\vert^2
\nonumber \\
&&+\xi\left\vert\varphi^{\{N,\xi\}}_{\rm L}(\mathbf{r})\right\vert^2
,
\end{eqnarray}
where L ($i=N_+$) and H ($i=N$) refer to the LUMO and HOMO 
of the $N$-electron KS system, respectively.  
By inserting the latter density   
into Eq.~(\ref{eq:GACE_integrand}) and taking $\alpha=\xi$, we finally
deduce from Eq.~(\ref{eq:mono_selfcons}) the analog
of the GOK-DFT optical gap expression for the fundamental gap, 
\be\label{eq:fun_gap_analog_gokdft}
E^N_{\rm g}=\varepsilon^{\{N,\xi\}}_{\rm L}
- \varepsilon^{\{N,\xi\}}_{\rm H}+
\left.\dfrac{\partial E^{{\{N,\xi\}}}_{\rm
xc}[n]}{\partial\xi}\right|_{n=n_{\hat{\Gamma}_0^{\{N,\xi\}}}}
.\ee
This is the central result of this work. Note that, when $\xi=0$, the
famous formula of Perdew and Levy~\cite{perdew1983physical} is recovered with
a much more
explicit density-functional expression for the DD.\\

\subsection{Two-weight generalization of the theory}\label{2-weight_edft}

\subsubsection{Extending the Levy--Zahariev shift-in-potential procedure
to ensembles}
In order to establish a connection between $N$-centered
ensemble DFT and the standard formulation of the fundamental
gap problem in DFT (which relies on fractional electron numbers), we propose
in the following to extend the theory to $N$-centered
ensembles where the removal and the addition of an electron can be
controlled 
independently. For that purpose, we introduce the generalized {\it
two-weight} 
$N$-centered ensemble density matrix operator,
\begin{eqnarray}
\hat{\Gamma}^{\{N,\bm{\xi}\}} & = & 
\sum_{\nu=\pm}\xi_{\nu}\hat{\Gamma}^{N_\nu} 
+ \left[1 - \sum_{\nu=\pm}\xi_\nu\dfrac{N_\nu}{N}
\right] \hat{\Gamma}^N, 
\end{eqnarray}
where ${\bm\xi}\equiv 
%\left\{\xi_\nu\right\}_{\nu=\pm}
\left( \xi_- , \xi_+ \right)
$ and the convexity conditions
$\xi_-\geq 0$, $\xi_+\geq 0$, and 
$\xi_-N_-+\xi_+N_+\leq N$ are fulfilled. Note that, by construction,
the $N$-centered ensemble density associated to $\hat{\Gamma}^{\{N,\bm{\xi}\}}$ still integrates to $N$,
and the single-weight formulation of $N$-centered ensemble DFT discussed
previously is simply recovered when $\xi_-=\xi_+=\xi$. 
The ensemble energy now reads
\begin{eqnarray}\label{eq:ens_E_two_weights}
E_0^{\{N,{\bm\xi}\}} & = & 
\sum_{\nu=\pm}\xi_\nu E_0^{N_\nu} 
  + \left[1 - 
\sum_{\nu=\pm}\xi_\nu\dfrac{N_\nu}{N}
\right]E_0^{N}.
\end{eqnarray}
Interestingly, if we extend the Levy--Zahariev shift-in-potential procedure~\cite{levy2014ground} 
to $N$-centered ground-state
ensembles as follows [note that the superscripts $\xi$ in Eq.~(\ref{eq:mono_selfcons}) should now be replaced by 
${\bm\xi}$ in the generalized two-weight theory],
\begin{eqnarray}\label{eq:shiftC}
\varepsilon_i^{\{N,{\bm\xi}\}}\rightarrow
\tilde{\varepsilon}_i^{\{N,{\bm\xi}\}}=\varepsilon_i^{\{N,{\bm\xi}\}}+C^{\{N,{\bm
\xi}\}}\left[n_{\hat{\Gamma}_0^{\{N,{\bm\xi}\}}}\right],
\ee
where the density-functional shift reads
\be\label{eq:densfun_Levy_shift}
C^{\{N,\bm{\xi}\}}[n]=\dfrac{E^{\{N,\bm{\xi}\}}_{\rm Hxc}[n]-\int\ddroit{\bf r}\;
v^{{\{N,{\bm \xi}\}}}_{\rm Hxc}[n](\mathbf{r})
%\dfrac{\delta E_{\rm
%Hxc}^{\{N,\bm{\xi}\}}[n]}{\delta n({\bf r})}
n({\bf r})}{\int\ddroit{\bf r}\;n({\bf r})}
,
\end{eqnarray}
the $N$-centered ground-state ensemble energy can be written as
a simple weighted
sum of shifted KS orbital energies.
Indeed, according to Eq.~(\ref{eq:min_KS-eDFT}) [where $\xi$ is replaced by $\bm{\xi}$], the $N$-centered
ground-state ensemble energy can be written as follows, 
\be
E_0^{{\{N,{\bm\xi}\}}}&=&
{\rm
Tr}\left[
\hat{\Gamma}_{\rm
KS}^{\{N,{\bm\xi}\}}
\left(\hat{T}+\hat{V}_{\rm ext}\right)\right]
+
E^{\{N,\bm{\xi}\}}_{\rm
Hxc}\left[n_{\hat{\Gamma}_{\rm KS}^{\{N,{\bm\xi}\}}}\right]
\nonumber\\
&=&
{\rm
Tr}\left[
\hat{\Gamma}_{\rm
KS}^{\{N,{\bm\xi}\}}
\left(\hat{T}+\hat{V}^{\{N,{\bm\xi}\}}_{\rm KS}\right)\right]
+
E^{\{N,\bm{\xi}\}}_{\rm
Hxc}\left[n_{\hat{\Gamma}_{\rm KS}^{\{N,{\bm\xi}\}}}\right]
\nonumber
\\
&&-
\int d{\bf r}\;
v^{{\{N,{\bm \xi}\}}}_{\rm Hxc}\left[n_{\hat{\Gamma}_{0}^{\{N,{\bm\xi}\}}}\right](\mathbf{r})
\;
n_{\hat{\Gamma}_{\rm KS}^{\{N,{\bm\xi}\}}}(\mathbf{r})
,
\ee
where 
\be\label{eq:KS_pot_Nc}
\hat{V}_{\rm KS}^{\{N,{\bm\xi}\}}=
\int d{\bf r}\;
v^{{\{N,{\bm \xi}\}}}_{\rm
KS}
(\mathbf{r})\,\hat{n}(\mathbf{r})
\ee 
and $n_{\hat{\Gamma}_{\rm KS}^{\{N,{\bm\xi}\}}}({\bf r})={\rm
Tr}\left[\hat{\Gamma}_{\rm KS}^{\{N,{\bm\xi}\}}\hat{n}({\bf r})\right]$.
Since the two densities $n_{\hat{\Gamma}_{\rm KS}^{\{N,{\bm\xi}\}}}$ and
$n_{\hat{\Gamma}_0^{\{N,{\bm\xi}\}}}$ are equal and integrate to $N$, we
obtain from Eqs.~(\ref{eq:mono_selfcons})
and~(\ref{eq:densfun_Levy_shift}),
\be\label{eq:shifting_pot}
E_0^{{\{N,{\bm\xi}\}}}&=&\sum_{\nu=\pm}\xi_\nu\sum^{N_\nu}_{i=1}\varepsilon_i^{\{N,{\bm\xi}\}}
%\nonumber\\
+\left[1-\sum_{\nu=\pm}\frac{\xi_\nu
N_\nu}{N}\right]\sum^{N}_{i=1}\varepsilon_i^{\{N,{\bm\xi}\}}
\nonumber\\
&&
+NC^{\{N,{\bm
\xi}\}}\left[n_{\hat{\Gamma}_0^{\{N,{\bm\xi}\}}}\right].
\ee 
Finally,
by 
rewriting
the last term in the right-hand side of Eq.~(\ref{eq:shifting_pot}) as follows,
\be
&&NC^{\{N,{\bm
\xi}\}}\left[n_{\hat{\Gamma}_0^{\{N,{\bm\xi}\}}}\right]=\sum_{\nu=\pm}\xi_\nu\sum^{N_\nu}_{i=1}C^{\{N,{\bm
\xi}\}}\left[n_{\hat{\Gamma}_0^{\{N,{\bm\xi}\}}}\right]
\nonumber\\
&&+\left(1-\sum_{\nu=\pm}\frac{\xi_\nu
N_\nu}{N}\right)\sum^{N}_{i=1}C^{\{N,{\bm
\xi}\}}\left[n_{\hat{\Gamma}_0^{\{N,{\bm\xi}\}}}\right],
\ee
and by using the definition of the shifted KS orbital energies in 
Eq.~(\ref{eq:shiftC}), we obtain the desired expression, 
\be\label{eq:ener_weighted_sum_epsi}
\hspace{-0.6cm}E_0^{\{N,{\bm\xi}\}}&=&
\left[1+\frac{\xi_--\xi_+}{N}\right]\sum^{N}_{i=1}\tilde{\varepsilon}_i^{\{N,{\bm\xi\}}}
-\xi_-\tilde{\varepsilon}_{\rm H}^{\{N,{\bm\xi}\}}
+\xi_+\tilde{\varepsilon}_{L}^{\{N,{\bm\xi}\}}.
\ee
In the non-interacting limit, it is readily seen from
Eq.~(\ref{eq:ener_weighted_sum_epsi}) that, unless $\xi_-=\xi_+=\xi$,
the $N$-centered ensemble does not describe a single-electron excitation
from the HOMO to the LUMO. In other words, the generalized two-parameter
$N$-centered ensemble non-interacting kinetic energy is not equal
anymore to its GOK-DFT analog. Interestingly, in the (very) particular
case $N=2$, the latter is actually recovered if the weight assigned to
the first excited state is set to $\xi_+$ (see Appendix~\ref{sec:Ts_proof}).\\

\subsubsection{Exact extraction of individual energies}

We will now show that, by using the shift-in-potential procedure
introduced previously and exploiting the linearity
in ${\bm{\xi}}$ 
of the ensemble energy, 
it becomes possible to extract 
individual $\mathcal{N}$-electron ground-state energies.
Starting from Eq.~(\ref{eq:ens_E_two_weights}) and noticing that
$E_0^{N}=E_0^{\{N,{\bm\xi}=0\}}$, we can express the exact $N$-electron
energy in terms of $E_0^{\{N,{\bm\xi}\}}$ and its derivatives as follows,
\begin{eqnarray}\label{eq:EN_from_ensE}
E_0^{N}=E_0^{\{N,{\bm\xi}\}}
-\sum_{\nu=\pm}\xi_\nu\dfrac{\partial E_0^{\{N,{\bm\xi}\}}}{\partial
\xi_\nu}
.
\end{eqnarray}
Moreover, as readily seen from Eq.~(\ref{eq:ens_E_two_weights}), the
$N_+$-
and $N_-$-electron energies can be extracted separately from the ensemble energy as
follows,  
\be\label{eq:E_pm_exp}
E^{N_\pm}_0=\dfrac{N_\pm}{N}E_0^{N}+\dfrac{\partial
E_0^{\{N,{\bm\xi}\}}}{\partial \xi_\pm}
.
\ee 
Note that, for convenience, Eqs.~(\ref{eq:EN_from_ensE}) and
(\ref{eq:E_pm_exp}) will be
compacted into a single equation,
\be\label{eq:allEN_from_ensE}
&&
\hspace{-0.4cm}
E_0^{\mathcal{N}}=\dfrac{\mathcal{N}}{N}E_0^{\{N,{\bm\xi}\}}
\nonumber\\
&&
\hspace{-0.4cm}
+\sum_{\nu=\pm}
\left[
\frac{(\mathcal{N}-N)(\mathcal{N}-N_{-\nu})}{2}
-\dfrac{\mathcal{N}\xi_\nu}{N}
\right]\dfrac{\partial E_0^{\{N,{\bm\xi}\}}}{\partial
\xi_\nu},
\ee
where $\mathcal{N}\in\{N_-,N,N_+\}$.
\\
%\\
%Let us now focus on the ensemble energy derivatives. 

Applying the
Hellmann--Feynman theorem to the variational ensemble energy
expression in Eq.~(\ref{eq:min_KS-eDFT}) [with the substitution
$\xi\rightarrow{\bm\xi}$] gives 
\be\label{eq:ensener_deriv_xi_pm}
\dfrac{\partial E_0^{\{N,{\bm\xi}\}}}{\partial
\xi_\pm}&=&{\rm
Tr}\left[
\left[\partial_{\xi_\pm}\hat{\Gamma}_{\rm
KS}^{\{N,{\bm\xi}\}}\right]\left(\hat{T}+\hat{V}^{\{N,{\bm\xi}\}}_{\rm KS}\right)\right]
\nonumber\\
&&+
\left.\dfrac{\partial E^{{\{N,\bm{\xi}\}}}_{\rm
xc}[n]}{\partial\xi_\pm}\right|_{n=n_{\hat{\Gamma}_0^{\{N,{\bm\xi}\}}}}
,
\ee
where 
\be\label{eq:deriv_densmat_op}
\partial_{\xi_\pm}\hat{\Gamma}_{\rm
KS}^{\{N,{\bm\xi}\}}=\left\vert\Phi_{N_\pm}^{\{N,{\bm\xi}\}}\right\rangle\left\langle\Phi_{N_\pm}^{\{N,{\bm\xi}\}}\right\vert
-\dfrac{N_\pm}{N}\left\vert\Phi_{N}^{\{N,{\bm\xi}\}}\right\rangle\left\langle\Phi_{N}^{\{N,{\bm\xi}\}}\right\vert, \nonumber \\
\ee
and the KS potential operator is defined in Eq.~(\ref{eq:KS_pot_Nc}).
Note that the $\mathcal{N}$-electron Slater determinants
$\Phi_{\mathcal{N}}^{\{N,{\bm\xi}\}}$ in Eq.~(\ref{eq:deriv_densmat_op})
are constructed from the KS orbitals
$\varphi_i^{\{N,{\bm\xi}\}}(\mathbf{r})$ in
Eq.~(\ref{eq:mono_selfcons}). Consequently,
Eq.~(\ref{eq:ensener_deriv_xi_pm}) can be simplified as follows,
\be\label{eq:dEens_dxi_pm_orb_ener}
\dfrac{\partial E_0^{\{N,{\bm\xi}\}}}{\partial
\xi_\pm}&=&
\pm\dfrac{1}{N}\sum^N_{i=1}\left(
\varepsilon_{N+\frac{1}{2}\pm\frac{1}{2}}^{\{N,{\bm\xi}\}}
-
\varepsilon_i^{\{N,{\bm\xi}\}}\right)
\nonumber\\
&&+
\left.\dfrac{\partial E^{{\{N,\bm{\xi}\}}}_{\rm
xc}[n]}{\partial\xi_\pm}\right|_{n=n_{\hat{\Gamma}_0^{\{N,{\bm\xi}\}}}}
.\ee
Since the shift introduced in Eq.~(\ref{eq:shiftC}) does not affect KS orbital energy differences,
\be
\varepsilon_j^{\{N,{\bm\xi}\}}
-
\varepsilon_i^{\{N,{\bm\xi}\}}
=
\tilde{\varepsilon}_j^{\{N,{\bm\xi}\}}
-
\tilde{\varepsilon}_i^{\{N,{\bm\xi}\}}
,
\ee
we finally deduce from  
Eqs.~(\ref{eq:ener_weighted_sum_epsi}),
(\ref{eq:allEN_from_ensE}), and (\ref{eq:dEens_dxi_pm_orb_ener})
the following exact expressions, 
\be\label{eq:ind_ener_from_N-eDFT}
E_0^{\mathcal{N}}=
\sum^{\mathcal{N}}_{i=1}\tilde{\varepsilon}_i^{\{N,{\bm\xi}\}}
+\sum_{\nu=\pm}
&&\left[
\frac{(\mathcal{N}-N)(\mathcal{N}-N_{-\nu})}{2}
-\dfrac{\mathcal{N}\xi_\nu}{N}
\right]
\nonumber\\
&&\times\left.\dfrac{\partial E^{{\{N,\bm{\xi}\}}}_{\rm
xc}[n]}{\partial\xi_\nu}\right|_{n=n_{\hat{\Gamma}_0^{\{N,{\bm\xi}\}}}}
.
\ee
%where $\mathcal{N}\in\{N_-,N,N_+\}$.
Eq.~(\ref{eq:ind_ener_from_N-eDFT}) is the second key result of this
work. As a direct consequence, the ionization potential (IP), denoted
$I^N$, and the
electron affinity (EA), denoted $A^N=I^{N_+}$, can now be extracted, in principle exactly, as
follows,
\be\label{eq:exact_IP_exp_compact}
&&
I^{N+\frac{1}{2}\pm\frac{1}{2}}=\pm\left(E^N-E^{N_\pm}\right)
=-\tilde{\varepsilon}_{N+\frac{1}{2}\pm\frac{1}{2}}^{\{N,{\bm\xi}\}}
\nonumber\\
&&
+
\sum_{\nu=\pm}
\left(\dfrac{\xi_\nu}{N}
+\frac{
N_{-\nu}
-N_\pm
}{2}\right)
\left.\dfrac{\partial E^{{\{N,\bm{\xi}\}}}_{\rm
xc}[n]}{\partial\xi_\nu}\right|_{n=n_{\hat{\Gamma}_0^{\{N,{\bm\xi}\}}}}
.
\ee
\\

As readily seen from Eq.~(\ref{eq:ind_ener_from_N-eDFT}), individual state properties can be extracted
exactly from the ensemble density. There is in principle no need to use
individual state densities for that purpose. Nevertheless, in practice,
it might be convenient to construct $N$-centered ground-state ensemble
xc density-functional approximations using individual densities, in the
spirit of the ensemble-based approach of Kraisler and
Kronik~\cite{kraisler2013piecewise}. Since the individual densities
are implicit functionals of the ensemble density, an optimized
effective potential would be needed. A similar strategy would apply if 
we want to remove ghost-interaction-type
errors~\cite{ensemble_ghost_interaction} by using an $N$-centered ensemble exact exchange
(EEXX) energy.\\

Finally, if we consider the conventional $N$-electron ground-state KS-DFT
limit of Eq.~(\ref{eq:ind_ener_from_N-eDFT}), i.e. ${\bm
\xi}=0$,
we recover 
the Levy--Zahariev expression
$E_0^{N}=\sum^{N}_{i=1}\tilde{\varepsilon}_i^{\{N,{\bm\xi}=0\}}$~\cite{levy2014ground}
for the $N$-electron energy and, in addition, we obtain the following
compact expressions for the anionic and cationic energies,
\be\label{eq:ENpm_xi_zero}
E^{N_\pm}_0=\sum^{N_\pm}_{i=1}\left(\tilde{\varepsilon}_i^{\{N,{\bm\xi}=0\}}
+ \frac{1}{N_\pm}\left.\dfrac{\partial E^{{\{N,\bm{\xi}\}}}_{\rm
xc}[n_{\Psi^N_0}]}{\partial\xi_\pm}\right|_{\substack{{\bm
\xi}=0}}\right),
\ee
where $n_{\Psi^N_0}$ denotes the exact $N$-electron ground-state
density. As well known and now readily seen from Eq.~(\ref{eq:ENpm_xi_zero}), it is 
impossible to describe all $\mathcal{N}$-electron ground-state energies
with the {\it same} potential. When an electron is added ($+$)/removed
($-$) 
to/from an $N$-electron system, an additional shift (second term in the
right-hand side of Eq.~(\ref{eq:ENpm_xi_zero})) 
is applied
to the already shifted KS orbital energies. 
Interestingly, we also recover
from Eq.~(\ref{eq:exact_IP_exp_compact}) a
more explicit form of the Levy--Zahariev IP expression~\cite{levy2014ground},
\begin{eqnarray}\label{eq:exact_IP_exp_xi_zero}
I^N&=&
-\tilde{\varepsilon}^{\{N,{\bm
\xi}=0\}}_{\rm H}
+
\left.\dfrac{\partial E^{{\{N,\bm{\xi}\}}}_{\rm
xc}[n_{\Psi^N_0}]}{\partial\xi_-}\right|_{\substack{{\bm
\xi}=0}}
,
\end{eqnarray}
where the second term in the right-hand side can be interpreted as the shifted Hxc
potential at position ${\bf
r}\rightarrow\infty$~\cite{levy2014ground}.\\

\begin{figure}
\resizebox{0.49\textwidth}{!}{
\includegraphics[scale=1]{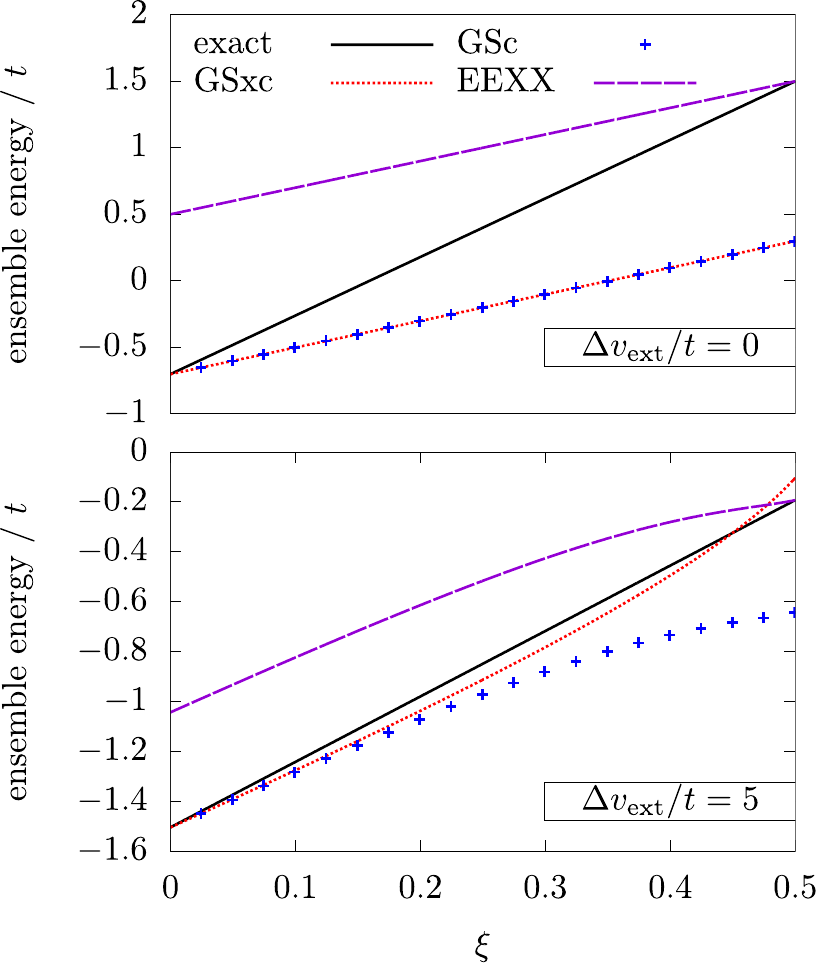}
}
\caption{(Color online) Exact and approximate $N$-centered ground-state ensemble
energies plotted as a function of $\xi$ 
for $N=2$ and $U/t=5$ in the symmetric (top panel) and 
asymmetric (bottom panel) Hubbard dimers. 
Approximate density-functional
energies are computed with the {\it exact} ensemble density.
The approximations in the
$N$-centered ensemble xc functional are $\mbox{EEXX:}~E^{\{N,\xi\}}_{\rm xc}(n)\approx E^{\{N,\xi\}}_{\rm x}(n)$,
$\mbox{GSxc:}~E^{\{N,\xi\}}_{\rm xc}(n)\approx E^{\{N,\xi=0\}}_{\rm
x}(n)+E^{\{N,\xi=0\}}_{\rm c}(n)$, and 
$\mbox{GSc:}~E^{\{N,\xi\}}_{\rm xc}(n)\approx E^{\{N,\xi\}}_{\rm
x}(n)+E^{\{N,\xi=0\}}_{\rm c}(n)$.
} 
\label{fig:Eens}
\end{figure}

\section{Application to the asymmetric Hubbard
dimer}\label{sec:Hubbard_dimer}

As a proof of concept, we apply in the following $N$-centered ensemble DFT to the asymmetric Hubbard
dimer~\cite{carrascal2015hubbard,carrascal2016corrigendum,deur2017exact}.
Despite its simplicity, the model is nontrivial and has become in recent
years a lab for analyzing and understanding failures of DFT or TD-DFT but also for
exploring new
ideas~\cite{li2018density,carrascal2018linear,deur2018exploring,sagredo2018can}.
By using such a model 
we also illustrate the fact that the theory applies not only to
exact {\it ab initio} Hamiltonians but also to lattice ones, which
might be of interest for modeling extended systems. In the Hubbard dimer, the
Hamiltonian is simplified as follows [we write operators in second
quantization],
\be\label{eq:Hamil_Hubbard_dimer_model}
\hat{T} &\rightarrow& -t
\sum_{\sigma=\uparrow\downarrow}(\hat{c}^\dagger_{0\sigma}\hat{c}_{1\sigma} + 
\hat{c}^\dagger_{1\sigma}\hat{c}_{0\sigma}),\hspace{0.2cm} \hat{W}_{\rm ee}\rightarrow
U\sum^1_{i=0}\hat{n}_{i\uparrow}\hat{n}_{i\downarrow}, 
\nonumber\\
\hat{V}_{\rm ext}&\rightarrow&\Delta v_{\rm ext}(\hat{n}_1 -
\hat{n}_0)/2,\hspace{0.3cm}\hat{n}_{i\sigma}=\hat{c}^\dagger_{i\sigma}\hat{c}_{i\sigma},
\ee
where $\hat{n}_i=\sum_{\sigma=\uparrow\downarrow}\hat{n}_{i\sigma}$ is
the density operator on site $i$ ($i=0,1$). Note that the external potential reduces to a single number $\Delta v_{\rm
ext}$ which controls the asymmetry of the model. 
The density also reduces 
to a single number $n=n_0$ which is the occupation of site 0 given that
$n_1 = N-n$. In the following, the central number of
electrons will be set to $N = 2$ so that the convexity condition reads
$\xi_+\leq (2-\xi_-)/3$. As shown in Appendix~\ref{sec:dimer}, the $N$-centered non-interacting kinetic and
EEXX energies can be expressed analytically as follows,  
\begin{eqnarray}\label{eq:Ts_ana_AND_eexx}
T_{\rm s}^{\{N,{\bm \xi}\}}(n) &=& - 2t \sqrt{(\xi_+ - 1)^2 - (n - 1)^2},
\nonumber\\
E_{\rm x}^{\{N,\bm\xi\}}(n)&  = & \dfrac{U}{2}\left[1 +\dfrac{\xi_+-\xi_-}{2}
+\left(1 - \dfrac{3\xi_++\xi_-}{2}\right) \left( \dfrac{n-1}{\xi_+ - 1}\right)^2\right]
\nonumber \\
&&-E_{\rm H}(n),
\end{eqnarray}
where the Hartree energy reads $E_{\rm H}(n)=
U\left(1 + (n - 1)^2\right)$. 
On the other hand, the correlation
energy can be computed exactly by   
Lieb maximization (see Ref.~\cite{deur2017exact} as well as Appendix~\ref{sec:dimer}). As readily seen from
Eq.~(\ref{eq:Ts_ana_AND_eexx}), an $N$-centered ensemble density $n$
is non-interacting $v$-representable if 
$| n- 1 | \leq 1 - \xi_+$. All the calculations have been performed with
$t=1$.\\

In Fig.~\ref{fig:Eens}, the total $N$-centered ground-state ensemble
energy is plotted as a function of $\xi_- = \xi_+ = \xi$. The exact
ensemble energy is linear in $\xi$, as it should.
Results obtained with various density-functional approximations are also
shown. In the simplest one, referred to as ground-state xc (GSxc), the
weight dependence is taken into account in neither the exchange nor
the correlation energies. In other words, the ``bare'' $\xi$-independent $N$-electron
ground-state xc functional is employed. On the other hand, both
EEXX-only (simply called EEXX) and the approximation
referred to as GSc take the weight dependence into account exactly in
the exchange energy. They differ only by the density-functional
correlation energy taken at $\xi=0$. The accurate parameterization of Carrascal
{\it et al.}~\cite{carrascal2015hubbard,carrascal2016corrigendum} has
been  
used for computing the latter correlation energy. Note that, as shown in
Appendix~\ref{sec:dimer}, the exact $N$-centered ensemble correlation
functional 
equals zero when $\xi=1/2$, thus making EEXX truly exact for this particular
weight. Returning to Fig.~\ref{fig:Eens}, we see that, in the symmetric case
[top panel], the approximate ensemble energies exhibit the expected
linear behavior in $\xi$. This is simply due to the fact that, in this case,
the ensemble density equals 1 and therefore,
as readily seen from Eq.~(\ref{eq:Ts_ana_AND_eexx}), 
all  
energy contributions vary (individually) linearly
with the ensemble weight. Moreover, for $n=1$, the exact ensemble exchange
energy is $\xi$-independent, since we consider the particular case $\xi_+=\xi_-$, which explains
why GSxc and GSc ensemble energies are on top of each other.     
We clearly see, when comparing GSc with the exact result, that the
correct ensemble energy slope, and therefore the proper description of the fundamental
gap, is recovered only when the weight
dependence is taken into account in the correlation
energy. This becomes even more critical in the asymmetric case [bottom
panel of Fig.~\ref{fig:Eens}] where approximations in the xc energy induce curvature, thus
leading to a weight-dependent fundamental gap, which is of course
unphysical.\\
\\
%%%%%%% FIGURE %%%%%%%%%%%%%%%%%%%
\begin{figure}
\resizebox{0.49\textwidth}{!}{
\includegraphics[scale=1.7]{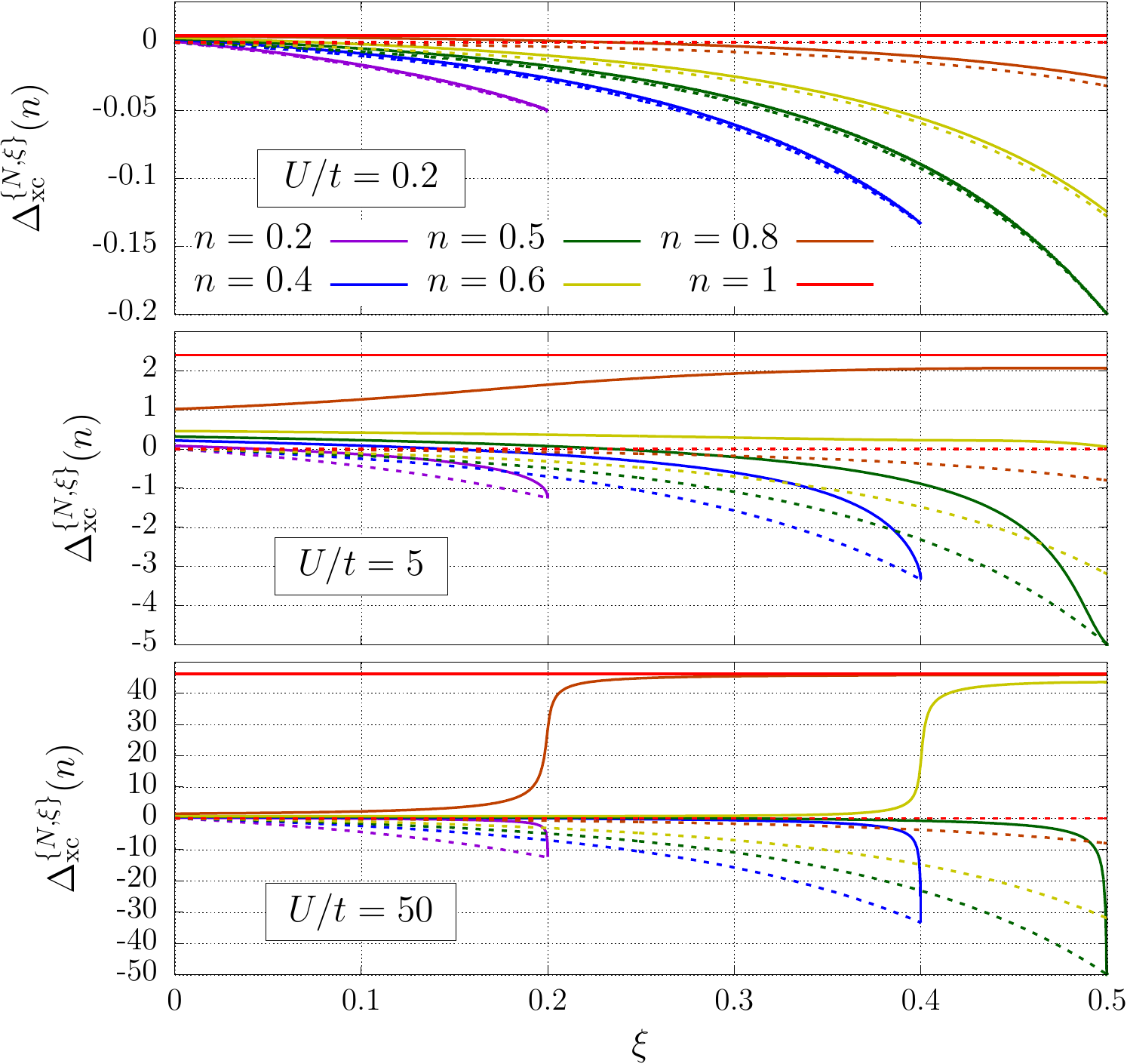}
}
\caption{(Color online) Exact GACE integrand plotted as a function of $\xi$ for
non-interacting $v$-representable densities [i.e. $\vert n-1\vert\leq
1-\xi$, thus leading to $0\leq \xi\leq {\rm inf}\{n,2-n,1/2\}$] and
$U/t=0.2$ (top panel), $U/t=5$ (middle panel), and $U/t=50$ (bottom panel). 
Full lines: xc integrand, dashed lines: exchange-only integrand [see
Eq.~(\ref{eq:EEXX-only_integrand})].
In each panel and for each integrand (xc or x-only), the curves are
ordered by decreasing density (the uppermost curve corresponds to $n=1$). See text for further details.}
\label{fig:integrand}
\end{figure}
\begin{figure}
\resizebox{0.49\textwidth}{!}{
\includegraphics[scale=1.7]{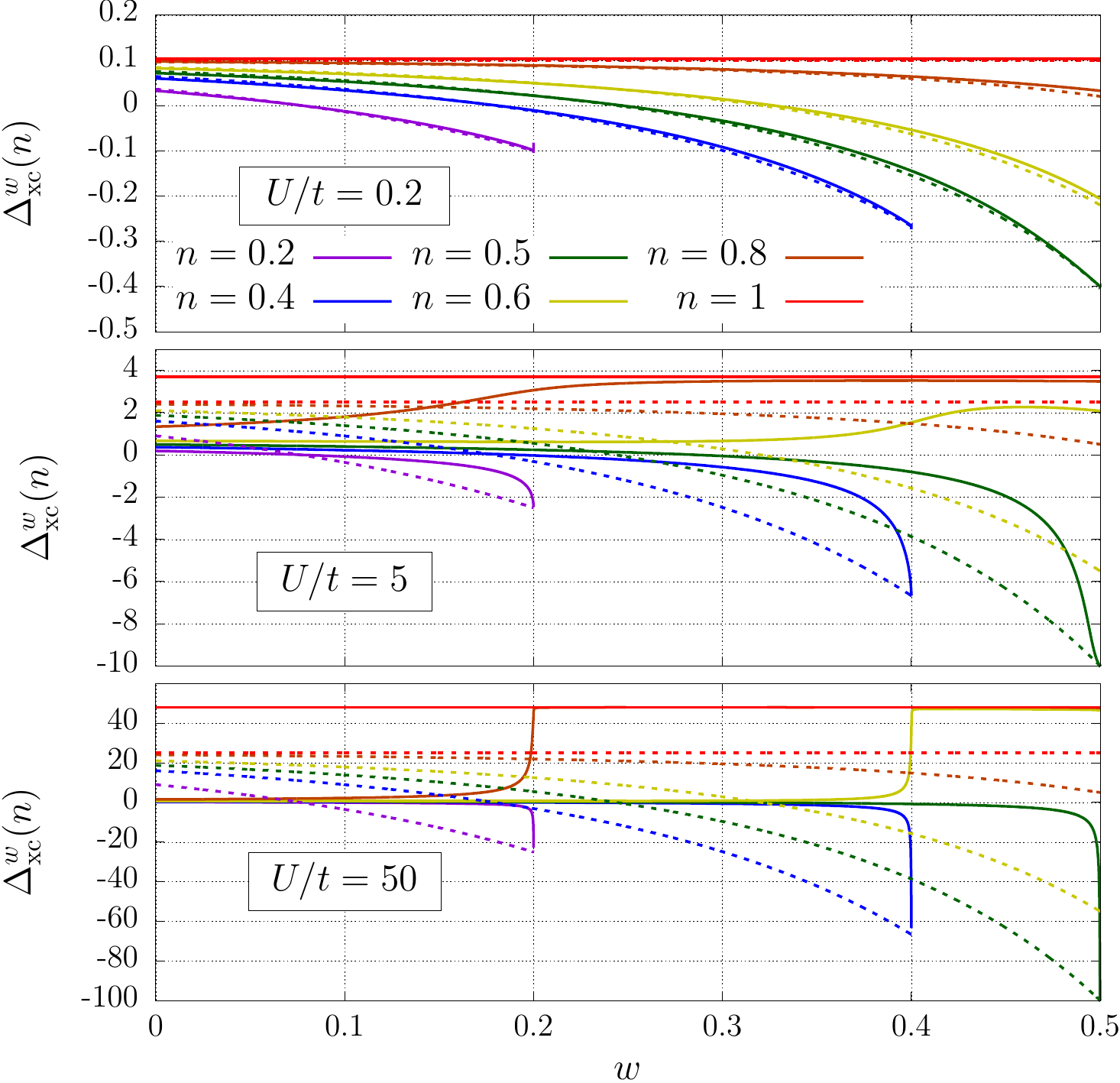}
}
\caption{(Color online) Same as Fig.~\ref{fig:integrand} but for the
GACE
integrand in GOK-DFT where $w$ is the weight assigned to the first (singlet)
two-electron excited state. See Ref.~\cite{deur2017exact} for further
details.}
\label{fig:integrand_GOKDFT}
\end{figure}
%%%%%%%%%%%%%%%%%%%%%%%%%%%%%%%
More insight into the weight dependence of the ensemble xc energy is given by the GACE
integrand in Eq.~(\ref{eq:GACE_integrand}) [see also Appendix~\ref{sec:dimer}].
As clearly seen when comparing Figs.~\ref{fig:integrand} and~\ref{fig:integrand_GOKDFT}, even though the integrand differs from its
GOK-DFT analog, they both vary similarly with the ensemble weight, in particular in the
strongly correlated regime. This can be rationalized by 
showing, in complete analogy with GOK-DFT (see Sec.~3.3 in
Ref.~\cite{deur2018exploring}), that
\be
E^{\{N,\xi\}}_{\rm xc}(n)\underset{U/t\rightarrow+\infty}{\longrightarrow}
U\times{\rm sup}\big\{\vert 1-n\vert,\xi\big\}-E_{\rm H}(n),
\ee
which gives (in the $U/t\rightarrow+\infty$ limit) 
$\partial E^{\{N,\xi\}}_{\rm
xc}(n)/\partial\xi=0$ in the non-interacting $v$-representable range
$0\leq \xi\leq n$
 if $0\leq n \leq 0.5$. For densities in the
range $0.5 < n\leq 1$,    
$\partial E^{\{N,\xi\}}_{\rm
xc}(n)/\partial\xi=U$ when $0.5\geq \xi > (1-n)$ and 
$\partial E^{\{N,\xi\}}_{\rm
xc}(n)/\partial\xi=0$ when $0\leq \xi < (1-n)$. The same analysis actually
holds for the GOK-DFT integrand~\cite{deur2018exploring} (see also
Ref.~\cite{deur2017exact} for further details). Let us finally mention that, in
the Lieb maximizations used to produce 
Figs.~\ref{fig:integrand} and~\ref{fig:integrand_GOKDFT}, both 
interacting {\it and} non-interacting potentials have been
determined 
numerically. In other words, we computed the expression in Eq.~(\ref{eq:general_gace_integrand_dimer})
rather than the one in Eq.~(\ref{eq:analytical_gace_integrand_dimer})
where the analytical expression for the KS potential is used. With such
a balanced description of both interacting and non-interacting gaps we
do not observe discontinuities in the GACE integrand at $n=\xi$ for densities in the
range $0\leq n \leq 0.5$ and large $U/t$ values, unlike in Fig.~6 of Ref.~\cite{deur2017exact}.
\\

\begin{figure}
\resizebox{0.49\textwidth}{!}{
\includegraphics[scale=1]{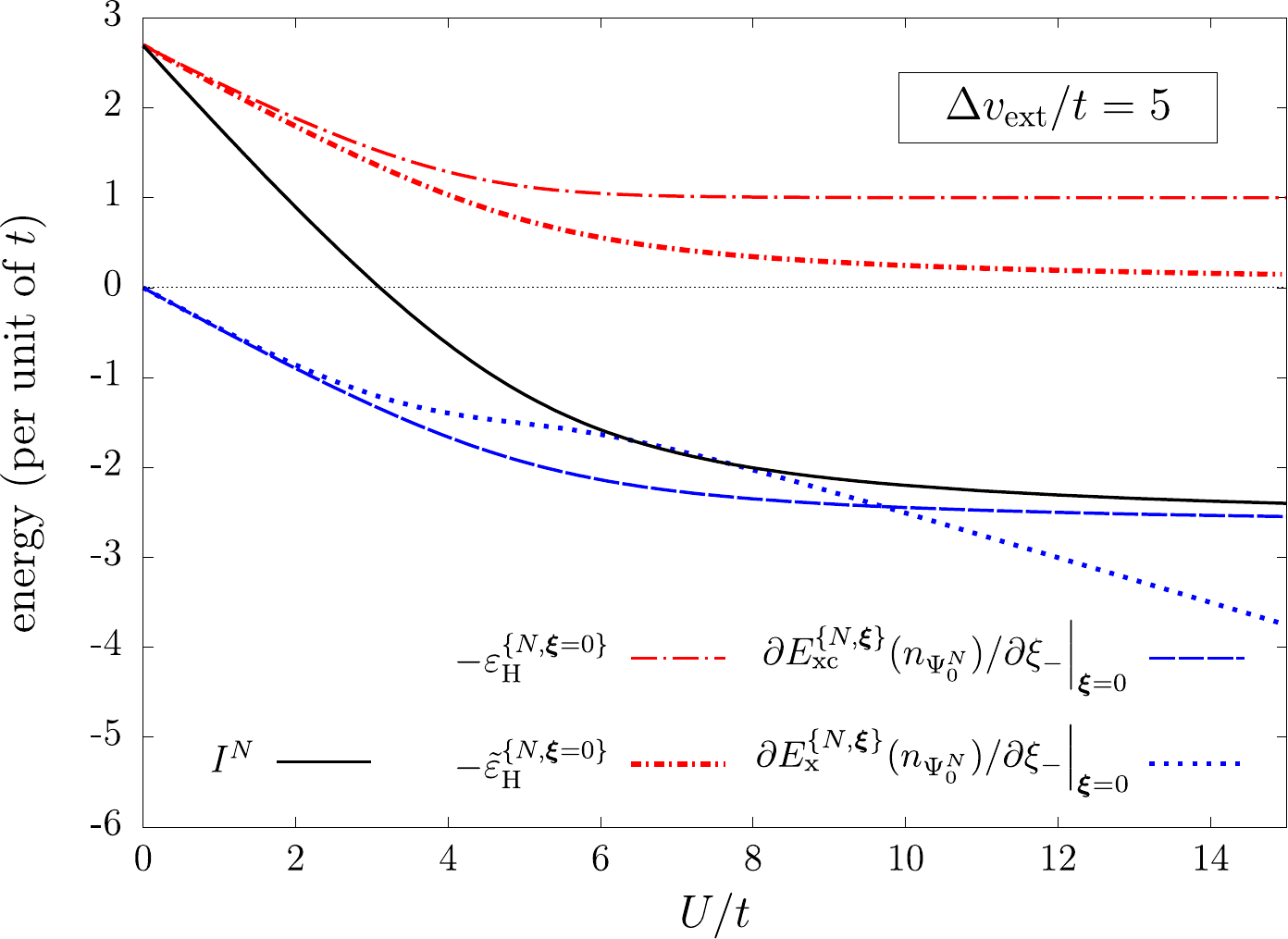}
}
\caption{(Color online) Contributions to the exact IP expression in
Eq.~(\ref{eq:exact_IP_exp_xi_zero}) plotted as a function of $U/t$ for
$\Delta v_{\rm ext}/t= 5$. The unshifted KS HOMO energy $\varepsilon^{\{N,{\bm
\xi}=0\}}_{\rm H}$ as well as the exchange-only contribution to the DD
are shown for analysis purposes. See text for further details.}
\label{fig:IP}
\end{figure}

Turning to the calculation of the IP,
Eq.~(\ref{eq:exact_IP_exp_xi_zero}) was verified by calculating
each (density-functional) contribution separately
(see Appendix~\ref{sec:dimer} for further details). Results
obtained for the asymmetric dimer are shown in
Fig.~\ref{fig:IP}.
As soon as the on-site repulsion is switched on (and up to
$U/t\approx4$), both the shifted KS HOMO
energy and the DD (second term in the right-hand side of
Eq.~(\ref{eq:exact_IP_exp_xi_zero})) contribute substantially to the IP.
Interestingly, in this regime of correlation and density, the shift-in-potential
procedure 
is not crucial. The unshifted KS HOMO energy, which is the analog for
the Hubbard dimer of the HOMO energy in a conventional KS-DFT
calculation, varies with $U$ through the
density. Note that the situation would be completely different in the
symmetric case [not shown] where $I^N(\Delta v_{\rm
ext}=0)=-t-E_0^N(\Delta v_{\rm ext}=0)$ and the unshifted KS HOMO energy equals
$-t$. By construction, the latter energy becomes 
$\frac{1}{2}E_0^{N}(\Delta v_{\rm ext}=0)$ 
 [which is $U$-dependent]
after shifting. As a result, in the symmetric case,
the shift and the DD equally contribute [by
$-(\frac{1}{2}E_0^{N}(\Delta v_{\rm ext}=0)+t)$] to the IP. 
Let us stress that, unlike the shifted KS orbital energies, the
unshifted ones are {\it not} uniquely defined. They are defined
up to a constant, like the KS potential, simply because the density
always integrates to a fixed integral number of electrons in
$N$-centered ensemble DFT, exactly like in a conventional $N$-electron
KS-DFT calculation. The shift will fix the KS orbital energy levels
according to Eq.~(\ref{eq:ener_weighted_sum_epsi}), which is equivalent
to the Levy--Zahariev shift-in-potential procedure~\cite{levy2014ground} when ${\bm
\xi}=0$. In the Hubbard dimer, our value of the unshifted KS HOMO energy has been fixed by
choosing a potential whose values on site 0 and 1 sum up to zero [see the potential operator expression in
Eq.~(\ref{eq:Hamil_Hubbard_dimer_model})]. To conclude, as mentioned previously, shifting the KS orbital
energies might be, in some cases, as important as taking into account the DD in the
calculation of the IP.   
Returning to Fig.~\ref{fig:IP}, the IP reduces to
the xc DD in the strongly correlated regime ($U/t\geq 10$) or,
equivalently, the shifted KS HOMO energy becomes negligible. Note that,
as expected, taking into account the exchange contribution to the DD
only leads to a poor description of the
IP when $U/t$ becomes large, thus illustrating the importance of weight dependence in both exchange and correlation energies.

\section{Conclusions and perspectives}\label{sec:conclusions}

We have shown that the fundamental gap problem, which is
traditionally formulated in grand canonical ensemble DFT, can be recast into a
canonical problem where the xc functional becomes ensemble weight dependent.
As a remarkable result, modeling the infamous DD becomes equivalent to modeling 
the weight dependence, exactly like in the optical gap problem. This key result, which is depicted in
Eq.~(\ref{eq:fun_gap_analog_gokdft}), opens up 
a new paradigm in the development of density functional approximations
for gaps which are computationally much cheaper than conventional
time-dependent post-DFT treatments. A natural step forward would be to apply the
approach, for example, to a finite uniform
electron gas~\cite{PRL09_Loos_2e_hypersphere}, thus providing an {\it ab initio} local density-functional
approximation that incorporates DDs through its weight dependence.
Work is currently in progress in this direction. 

\begin{acknowledgments}
The authors thank the Ecole Doctorale des Sciences Chimiques 222
(Strasbourg) and the ANR (MCFUNEX project, Grant No. ANR-14-CE06-
0014-01) for funding.
\end{acknowledgments}

\appendix 
\section{Connection between the non-interacting kinetic energy functional in GOK-DFT and its analog in $N$-centered ensemble DFT.}\label{sec:Ts_proof}

For the sake of generality, we will first consider the generalized
(interacting) {\it two-weight} formulation
of the theory [which is introduced in Sec.~\ref{2-weight_edft}] and denote $E_0^{{\{N,{\bm \xi}\}}}[v]$ the $N$-centered ground-state
ensemble energy of $\hat{T}+\hat{W}_{\rm
ee}+\int\ddroit\mathbf{r}\;v(\mathbf{r})\hat{n}(\mathbf{r})$. According
to the variational principle in Eq.~(\ref{eq:ens_var_principle_dens_vext}), the
following inequality holds for {\it any} $N$-electron density $n$ and potential $v$,  
\be
E_0^{{\{N,{\bm \xi}\}}}[v]\leq 
F^{{\{N,{\bm \xi}\}}}[n]+
\int\ddroit{\mathbf{r}}\,v(\mathbf{r})n(\mathbf{r}),
\ee
or, equivalently,
\be
F^{{\{N,{\bm \xi}\}}}[n]
\geq 
E_0^{{\{N,{\bm \xi}\}}}[v]
-
\int\ddroit{\mathbf{r}}\,v(\mathbf{r})n(\mathbf{r})
,
\ee
thus leading to the Legendre-Fenchel transform-based expression,
\begin{eqnarray}\label{eq:F_LF_general}
F^{{\{N,{\bm \xi}\}}}[n]
= \sup_v \Big \lbrace 
E_0^{{\{N,{\bm \xi}\}}}[v]
-
\int\ddroit{\mathbf{r}}\,v(\mathbf{r})n(\mathbf{r})
 \Big \rbrace.
\end{eqnarray}
In the non-interacting case, Eq.~(\ref{eq:F_LF_general}) becomes
\begin{eqnarray}\label{eq:kinetic}
T_{\rm s}^{\{N,{\bm\xi}\}}[n] = \sup_v \Big \lbrace \mathcal{E}_{\rm
KS}^{\{N,{\bm\xi}\}}[v] - \int \ddroit \mathbf{r}
\, v(\mathbf{r})n(\mathbf{r}) \Big \rbrace,
\end{eqnarray}
where $\mathcal{E}_{\rm KS}^{\{N,{\bm\xi}\}}[v]$ is the $N$-centered ground-state ensemble energy
of $\hat{T} + \int \ddroit \mathbf{r}\;
v(\mathbf{r})\hat{n}(\mathbf{r})$. According to
Eq.~(\ref{eq:ens_E_two_weights}), the latter energy can be expressed as
follows in terms of the $v$-dependent orbital energies [i.e. the
eigenvalues of $-\frac{1}{2}\nabla_{\bf r}^2+v(\mathbf{r})$],   
\begin{eqnarray}
\label{eq:nonint_energy}
\mathcal{E}_{\rm KS}^{\{N,\bm{\xi}\}}[v] 
& = & \xi_- \sum_{i=1}^{N_-} \varepsilon_{i}[v] + \xi_+ \sum_{i=1}^{N_+} \varepsilon_{i}[v]
\nonumber \\
& &+ \left[1 - \dfrac{N_-\xi_-}{N}  -  \dfrac{N_+\xi_+}{N} \right] \sum_{i=1}^N
\varepsilon_i[v],
\end{eqnarray}
or, equivalently,
\begin{eqnarray}\label{eq:nonint_N-c_ener}
\mathcal{E}_{\rm KS}^{\{N,\bm{\xi}\}}[v] 
 &=& \left[1+\frac{\xi_--\xi_+}{N}\right]\sum^{N-2}_{i=1}\varepsilon_i[v]
\nonumber\\
&& +\left[1+\frac{\xi_--\xi_+}{N}\right]\varepsilon_{N_-}[v]
\nonumber\\
 &&+\left[1-\dfrac{N_-\xi_-+\xi_+}{N}\right]\varepsilon_{\rm H}[v]
\nonumber\\
&&+ \xi_+\varepsilon_{\rm L}[v], 
\end{eqnarray}
where H and L refer to the HOMO and the LUMO of the $N$-electron KS
system.\\

In GOK-DFT, the non-interacting ensemble kinetic energy functional
reads~\cite{deur2017exact,deur2018exploring} 
\begin{eqnarray}\label{eq:kinetic_gokdft}
T_{\rm s}^{N,w}[n] = \sup_v \Big \lbrace \mathcal{E}_{\rm
KS}^{N,w}[v] - \int \ddroit \mathbf{r}
v(\mathbf{r})n(\mathbf{r}) \Big \rbrace,
\end{eqnarray}
where the ensemble energy is obtained by averaging the $N$-electron ground-
and first-excited-state energies of $\hat{T} + \int \ddroit \mathbf{r}\;
v(\mathbf{r})\hat{n}(\mathbf{r})$ [$w$ is the weight assigned to the
excited state],
\begin{eqnarray}
\label{eq:nonint_energy_GOKDFT}
\mathcal{E}_{\rm KS}^{N,w}[v] & = & (1-w)\sum_{i=1}^{N} \varepsilon_i[v] 
+ w\left( \sum_{i=1}^{N_-} \varepsilon_i[v] +\varepsilon_{\rm L}[v]\right) \nonumber \\
 & = & \sum_{i=1}^{N_-} \varepsilon_i[v] + (1-w)\varepsilon_{\rm H}[v] +
w\varepsilon_{\rm L}[v].
\end{eqnarray}
As readily seen from Eqs.~(\ref{eq:nonint_N-c_ener}) and
~(\ref{eq:nonint_energy_GOKDFT}), in the particular case
${\bm \xi}=\underline{\xi}\equiv (\xi,\xi)$ [i.e. $\xi_-=\xi_+=\xi$], we have 
\be
\mathcal{E}_{\rm KS}^{\{N,\underline{\xi}\}}[v]=\mathcal{E}_{\rm
KS}^{N,w=\xi}[v], 
\ee  
thus leading, according to Eqs.~(\ref{eq:kinetic}) and (\ref{eq:kinetic_gokdft}), to the equality
\be
T_{\rm s}^{\{N,\xi\}}[n]:= T_{\rm
s}^{\{N,\underline{\xi}\}}[n]=T_{\rm s}^{N,w=\xi}[n].
\ee
Returning to the general (two-weight) expression in
Eq.~(\ref{eq:nonint_N-c_ener}), we note that, in the particular case
$N=2$, the first term in the right-hand side vanishes. Moreover, since
$\varepsilon_1[v] = \varepsilon_2[v] = \varepsilon_{\rm H}[v]$ within the
conventional spin-restricted formalism, it comes
\begin{eqnarray}
\label{eq:nonint_energy_N=2}
\mathcal{E}_{\rm KS}^{\{N=2,\bm{\xi}\}}[v] & = & 
(2 - \xi_+)\varepsilon_{\rm H}[v] + \xi_+ \varepsilon_{\rm L}[v]
\nonumber\\
&=&\mathcal{E}_{\rm KS}^{N=2,w=\xi_+}[v]
,\end{eqnarray}
thus leading to the equality
\be\label{eq:link_Ts_gok-dft_Nc}
T_{\rm s}^{\{N=2,{\bm\xi}\}}[n]=T_{\rm s}^{N=2,w=\xi_+}[n].
\ee

\section{Exact expression for the one-weight GACE integrand in $N$-centered ensemble DFT}\label{appendix:single_weight_GACE_int}

According to Eq.~(\ref{eq:GACE}), the GACE integrand reads 
\begin{eqnarray}
\Delta_{\rm xc}^{\{N,\alpha\}}[n] = \dfrac{\partial E_{\rm xc}^{\{N,\alpha\}}[n]}{\partial \alpha},
\end{eqnarray}
or, equivalently [see Eqs.~(\ref{eq:KS_decomp_Fens}) and (\ref{Nc-Hxc_equal_H_plus_xc})],
\begin{eqnarray}
\Delta_{\rm xc}^{\{N,\alpha\}}[n] 
=  
 \dfrac{\partial F^{\{N,\alpha\}}[n]}{\partial \alpha} -
 \dfrac{\partial T_{\rm s}^{\{N,\alpha\}}[n]}{\partial \alpha},
\end{eqnarray}
where, with the notations used in Appendix~\ref{sec:Ts_proof},
$F^{\{N,\alpha\}}[n]:=F^{\{N,\underline{\alpha}\}}[n]$, $T_{\rm
s}^{\{N,\alpha\}}[n]:=T_{\rm s}^{\{N,\underline{\alpha}\}}[n]$, and
$\underline{\alpha}=(\alpha,\alpha)$. If we
denote $v^{\{N,\underline{\alpha}\}}[n]$ and $v_{\rm KS}^{\{N,\underline{\alpha}\}}[n]$ 
the (stationary) maximizing potentials in Eqs.~(\ref{eq:F_LF_general}) and~(\ref{eq:kinetic})
[where ${\bm\xi}=\underline{\alpha}$],
respectively, it comes 
\begin{eqnarray}\label{eq:dF_dalpha}
\dfrac{\partial F^{\{N,\alpha\}}[n]}{\partial \alpha}
& = &\left.\dfrac{\partial E_0^{{\{N,\underline{\alpha}\}}}[v]}{\partial
\alpha}\right|_{v=v^{\{N,\underline{\alpha}\}}[n]} 
%E_g^{\{N,\alpha\}}[n],
\end{eqnarray}
and
\be
\dfrac{\partial T_{\rm s}^{\{N,\alpha\}}[n]}{\partial \alpha}=
\left.\dfrac{\partial 
\mathcal{E}_{\rm KS}^{\{N,\underline{\alpha}\}}[v]
}{\partial
\alpha}\right|_{v=v_{\rm KS}^{\{N,\underline{\alpha}\}}[n]}, 
\ee
where, according to Eq.~(\ref{eq:ens_E_two_weights}) with ${\bm\xi}=\underline{\alpha}$ [or, equivalently, Eq.~(\ref{eq:ens_ener})],
\be
\dfrac{\partial E_0^{{\{N,\underline{\alpha}\}}}[v]}{\partial
\alpha}=E^N_{\rm g}[v]
\ee
is the fundamental gap for the interacting $N$-electron system with Hamiltonian $\hat{T}+\hat{W}_{\rm
ee} + \int \ddroit \mathbf{r}\;
v(\mathbf{r})\hat{n}(\mathbf{r})$, and
\be
\dfrac{\partial
\mathcal{E}_{\rm KS}^{\{N,\underline{\alpha}\}}[v]
}{\partial
\alpha}=\varepsilon_{\rm L}[v]-\varepsilon_{\rm H}[v]
\ee
is the HOMO-LUMO gap for the $N$-electron non-interacting system with
Hamiltonian $\hat{T} + \int \ddroit \mathbf{r}\;
v(\mathbf{r})\hat{n}(\mathbf{r})$. 
Let us stress that, when the interacting and non-interacting potentials
are equal to $v^{\{N,\underline{\alpha}\}}[n]$ and $v_{\rm
KS}^{\{N,\underline{\alpha}\}}[n]$, respectively, both systems have the
same $N$-centered ground-state ensemble density
with weight
$\underline{\alpha}$, namely $n$. We finally recover Eq.~(\ref{eq:GACE_integrand}) by
using the following notations,
\be
E^{\{N,\alpha\}}_{\rm g}[n]&=&E^N_{\rm g}\left[v^{\{N,\underline{\alpha}\}}[n]\right]
\nonumber\\
\varepsilon^{\{N,\alpha\}}_{i}[n]&=&\varepsilon_{i}\left[v_{\rm
KS}^{\{N,\underline{\alpha}\}}[n]\right],\hspace{0.2cm} i={\rm H,L}. 
\ee

\section{Technical details about $N$-centered ensemble DFT for the asymmetric Hubbard dimer}\label{sec:dimer}

In the following the central number of electrons is set to $N=2$.

\subsection{Hole-particle symmetry}\label{sec:hole-particle}
In this section we explain how the 3-electron ground-state energy of the
Hubbard dimer can be trivially obtained from the one-electron one by
using hole-particle symmetry. If we apply the following hole-particle
transformation to the annihilation operators in
Eq.~(\ref{eq:Hamil_Hubbard_dimer_model}),  
\begin{eqnarray}
\hat{c}_{0\sigma}&\rightarrow&
\hat{b}_{0\sigma}=\hat{c}^\dagger_{0\sigma},  \nonumber\\
\hat{c}_{1\sigma}&\rightarrow&
\hat{b}_{1\sigma}=-\hat{c}^\dagger_{1\sigma},
\end{eqnarray}
then the Hubbard dimer Hamiltonian, 
\be\label{eq:HD_Hamilt}
&&\hat{H}(\Delta v)= -t
\sum_{\sigma=\uparrow\downarrow}(\hat{c}^\dagger_{0\sigma}\hat{c}_{1\sigma} +
\hat{c}^\dagger_{1\sigma}\hat{c}_{0\sigma})
\nonumber\\
&&
+U\sum^1_{i=0} \hat{c}_{i\uparrow}^\dagger \hat{c}_{i\uparrow} \hat{c}_{i\downarrow}^\dagger              \hat{c}_{i\downarrow}
+\dfrac{\Delta v}{2}
\sum_{\sigma=\uparrow\downarrow}
\left(\hat{c}_{1\sigma}^\dagger\hat{c}_{1\sigma}-\hat{c}_{0\sigma}^\dagger\hat{c}_{0\sigma}\right)
,\nonumber
\\
\ee
can be rewritten as follows, according to the anti-commutation rules,
\be
&&\hat{H}(\Delta v)= -t
\sum_{\sigma=\uparrow\downarrow}(\hat{b}^\dagger_{0\sigma}\hat{b}_{1\sigma} +
\hat{b}^\dagger_{1\sigma}\hat{b}_{0\sigma})
\nonumber\\
&&
+U\sum^1_{i=0} \hat{b}_{i\uparrow} \hat{b}^\dagger_{i\uparrow} \hat{b}_{i\downarrow}
\hat{b}_{i\downarrow}
^\dagger
+\dfrac{\Delta v}{2}
\sum_{\sigma=\uparrow\downarrow}
\left(\hat{b}_{1\sigma}\hat{b}^\dagger_{1\sigma}-\hat{b}_{0\sigma}\hat{b}^\dagger_{0\sigma}\right)
,
\nonumber
\\
\ee
or, equivalently,
\be\label{eq:Ham_after_hp_transf}
&&\hat{H}(\Delta v)= -t
\sum_{\sigma=\uparrow\downarrow}(\hat{b}^\dagger_{0\sigma}\hat{b}_{1\sigma} +
\hat{b}^\dagger_{1\sigma}\hat{b}_{0\sigma})
\nonumber\\
&&
+U\sum^1_{i=0}
 \hat{b}^\dagger_{i\uparrow}
 \hat{b}_{i\uparrow}
\hat{b}_{i\downarrow}
^\dagger
 \hat{b}_{i\downarrow}
-
\dfrac{\Delta v}{2}
\sum_{\sigma=\uparrow\downarrow}
\left(
\hat{b}^\dagger_{1\sigma}
\hat{b}_{1\sigma}
-
\hat{b}^\dagger_{0\sigma}
\hat{b}_{0\sigma}
\right)
\nonumber\\
&&- U\sum^1_{i=0}\sum_{\sigma=\uparrow\downarrow} \hat{b}_{i\sigma}^\dagger\hat{b}_{i\sigma}
+2U.
\ee
As readily seen from Eqs.~(\ref{eq:HD_Hamilt}) and
(\ref{eq:Ham_after_hp_transf}), the $\mathcal{N}$-electron ground-state
energy $E^{\mathcal{N}}_0(\Delta v)$ of $\hat{H}(\Delta v)$ is connected
to the $(4-\mathcal{N})$-electron ground-state energy of
$\hat{H}(-\Delta v)$ as follows,
\be
E^{\mathcal{N}}_0(\Delta v)=E^{4-\mathcal{N}}_0(-\Delta
v)+U(\mathcal{N}-2).
\ee 
In the particular case $\mathcal{N}=3$, we obtain the useful result [let
us recall that $N=2$]
\be\label{eq:ener_3e_from1e}
E^{N_+}_0(\Delta v)=E^{N_-}_0(-\Delta v)+U.
\ee

\subsection{Exact functionals}\label{sec:exact_functionals}

%F,Ts,Eens,E0N,E0N_1,E0N_3,eens,e0N,e0N_1,e0N_3, deltav_KS^xi deltav^xi,
%nens_exact, n1,n2,n3, EHx^xi.
In the two-site Hubbard model,
%(see the Hamiltonian in
%Eq.~(\ref{eq:Hamil_Hubbard_dimer_model}))
the Legendre--Fenchel
transform in Eq.~(\ref{eq:F_LF_general}) can be rewritten as follows,
\begin{eqnarray}\label{eq:LF_F_hubbard}
F^{\{N,{\bm \xi}\}}(n)=\sup_{\Delta v}\Big\lbrace E^{\{N,{\bm \xi}\}}(\Delta v)- 
{\Delta v}(1 - n)
\Big\rbrace,
\end{eqnarray}
by analogy with GOK-DFT~\cite{deur2017exact}. The interacting ensemble
energy reads [with $N=2$], 
\begin{eqnarray}\label{eq:ensemble_energy_Hubbard}
E^{\{N,\bm{\xi}\}}(\Delta v) &= &  \xi_- E_0^{N_-}(\Delta v) + \xi_+ E_0^{N_+}(\Delta v) \nonumber \\
&&+ 
\left[1-\dfrac{\xi_-}{2}
- \dfrac{3\xi_+}{2} \right]E_0^{N}(\Delta v)
,
\end{eqnarray}
where the one-electron energy is simply the energy of the HOMO for the
non-interacting two-electron system~\cite{deur2017exact},
\begin{eqnarray}\label{eq:E0N-}
E_0^{N_-}(\Delta v)  =  \varepsilon_{\rm H}(\Delta v)= 
- \sqrt{t^2 + (\Delta v^2 / 4)}
,
\end{eqnarray}
and, according to Eq.~(\ref{eq:ener_3e_from1e}), the 3-electron energy
equals 
\begin{eqnarray}\label{eq:E0N+}
 E_0^{N_+}(\Delta v) = \varepsilon_{\rm H}(\Delta v) + U.
\end{eqnarray}
Therefore, Eq.~(\ref{eq:LF_F_hubbard}) can be rewritten as follows,
\begin{eqnarray}\label{eq:LF_F_hubbard_simplified}
&&F^{\{N,{\bm \xi}\}}(n)=\sup_{\Delta v}\Bigg\lbrace
(\xi_-+\xi_+)\varepsilon_{\rm H}(\Delta v)+\xi_+U
\nonumber\\
&&+
\left[1-\dfrac{\xi_-}{2}
- \dfrac{3\xi_+}{2} \right]E_0^{N}(\Delta v)
 - 
{\Delta v}(1 - n)
\Bigg\rbrace,
\end{eqnarray}
where the two-electron ground-state energy has the following analytical expression~\cite{carrascal2015hubbard,carrascal2016corrigendum}, 
\begin{eqnarray}
E_0^{N}(\Delta v)& = & \dfrac{4t}{3}\left(u-w\,{\rm sin}\left(\theta +\dfrac{\pi}{6}\right)\right)
\end{eqnarray}
with
\begin{eqnarray}
u& = & U/2t,
\\ 
w& = & \sqrt{3(1+\nu^2)+u^2},
\\
\nu & = & \Delta v/2t,
\end{eqnarray}
and
\begin{eqnarray}
{\rm cos}(3\theta)=\left(9(\nu^2-1/2)-u^2\right)u/w^3.
\end{eqnarray}
Note that the maximizing potential $\Delta v^{\{N,{\bm\xi}\}}(n)$ in
Eq.~(\ref{eq:LF_F_hubbard_simplified}), which fulfills the following stationarity
condition,
\be\label{eq:Vphys_dF_over_dn}
\dfrac{\partial F^{\{N,{\bm \xi}\}}(n)}{\partial n}=\Delta
v^{\{N,{\bm\xi}\}}(n),
\ee
has no simple analytical
expression. However, since the potential-functional quantity to be
maximized can be expressed analytically, it is straightforward to compute
the exact value of $\Delta v^{\{N,{\bm\xi}\}}(n)$ for any density $n$,
like in GOK-DFT~\cite{deur2017exact}.\\
The non-interacting $N$-centered ground-state ensemble kinetic energy
functional, i.e. the functional obtained from
Eq.~(\ref{eq:LF_F_hubbard_simplified}) when $U=0$, has a simple
analytical expression given in Eq.~(\ref{eq:Ts_ana_AND_eexx}). This is a
direct consequence of Eq.~(\ref{eq:link_Ts_gok-dft_Nc}) and Eq.~(57) in
Ref.~\cite{deur2017exact}. Note that, by considering
Eq.~(\ref{eq:Vphys_dF_over_dn}) in the particular case $U=0$, we can
express the KS potential analytically as follows, 
\begin{eqnarray}\label{eq:KS_pot}
\Delta v_{\rm KS}^{\{N,{\bm \xi}\}}(n) = \dfrac{2t(n-1)}{\sqrt{(\xi_+ - 1)^2 - (1 - n)^2}}.
\end{eqnarray}
Turning to the $N$-centered EEXX energy, let us start with the formal
expression
\be
E_{\rm x}^{\{N,\bm{\xi}\}}(n)=U\left[\left.\dfrac{\partial
F^{\{N,\bm{\xi}\}}(n)}{\partial U}\right|_{U=0}\right]
-E_{\rm H}(n),\ee 
where, according to Eq.~(\ref{eq:LF_F_hubbard_simplified}),
\be\label{eq:dF_over_dU_Nc}
&&
\left.\dfrac{\partial
F^{\{N,\bm{\xi}\}}(n)}{\partial U}
\right|_{U=0}
=\xi_+
\nonumber\\
&&+\left[1-\dfrac{\xi_-}{2}
- \dfrac{3\xi_+}{2} \right]\left.\dfrac{\partial E_0^{N}(\Delta
v)}{\partial U}\right|_{U=0,\Delta v=\Delta v_{\rm KS}^{\{N,{\bm \xi}\}}(n)}
,\ee
and, according to Eq.~(A7) in Ref.~\cite{deur2017exact}, 
\be\label{eq:dE_over_dU_express}
\left.\dfrac{\partial E_0^{N}(\Delta
v)}{\partial U}\right|_{U=0
}=\dfrac{4t^2-8\varepsilon^2_{\rm H}(\Delta v)}{4t^2+(\Delta v)^2-12\varepsilon^2_{\rm H}(\Delta v)}
,\ee
or, equivalently [see Eq.~(\ref{eq:E0N-})],
\be\label{eq:dE_over_dU_express2}
\left.\dfrac{\partial E_0^{N}(\Delta
v)}{\partial U}\right|_{U=0
}=\dfrac{2t^2+(\Delta v)^2}{4t^2+(\Delta v)^2}
.\ee
Combining Eqs.~(\ref{eq:KS_pot}), and
(\ref{eq:dF_over_dU_Nc}) with Eq.~(\ref{eq:dE_over_dU_express2}) gives
\be
&&
\left.\dfrac{\partial
F^{\{N,\bm{\xi}\}}(n)}{\partial U}
\right|_{U=0}
=\xi_+
\nonumber\\
&&+\left[1-\dfrac{\xi_-}{2}
- \dfrac{3\xi_+}{2} \right]
\times\dfrac{(1-\xi_+)^2+(1-n)^2}{2(1-\xi_+)^2},
\ee
thus leading to the expression in Eq.~(\ref{eq:Ts_ana_AND_eexx}) for the EEXX functional. 

\subsection{Correlation energy at the border of the representability domain}

Let us consider the one-weight formulation of $N$-centered ensemble DFT
(i.e. $\xi_-=\xi_+=\xi$). We will show in the following that, at the
border of the non-interacting $v$-representability domain
[i.e. when $n=1\pm(1-\xi)$], the $N$-centered ground-state ensemble correlation
energy equals zero. The proof follows closely its analog in GOK-DFT (see
Appendix C in Ref.~\cite{deur2018exploring}).
\\

According to Eq.~(\ref{eq:LF_F_hubbard_simplified}), the (unique)
maximizing potential $\Delta v^{\{N,\xi\}}(n):=\Delta
v^{\{N,\underline{\xi}\}}(n)$ with $\underline{\xi}=(\xi,\xi)$ fulfills
the following stationarity condition,  
\be\label{eq:stationarity_cond_LF}
&&\left[-\dfrac{\xi\Delta v}{2\sqrt{t^2+(\Delta v^2/4)}}+(1-2\xi).\dfrac{\partial
E_0^N(\Delta v)}{\partial \Delta v}\right]_{\Delta v=\Delta v^{\{N,\xi\}}(n)}
\nonumber\\
&&=1-n.\ee
Since $E^N_0(\Delta v)=U-\vert\Delta v\vert$ and $
{\partial E^N_0(\Delta v)}/{\partial \Delta v}=-{\Delta
v}/{\vert\Delta v\vert}
$
when $\vert\Delta v\vert/t\rightarrow +\infty$ and $\vert\Delta
v\vert>U$ [see Ref.~\cite{deur2018exploring}],  
we conclude that the stationarity condition in
Eq.~(\ref{eq:stationarity_cond_LF}) is fulfilled for $n=1\pm (1-\xi)$ when $\vert\Delta
v\vert/t\rightarrow +\infty$ and $\Delta v/(n-1)$ is {\it positive}.
As a result, in this particular case, the Legendre--Fenchel transform in
Eq~(\ref{eq:LF_F_hubbard_simplified}) can be simplified as follows,
\be\label{eq:Fw_border}
&&
-\xi\vert\Delta v\vert+\xi U+(1-2\xi).\Big(U-\vert\Delta v\vert\Big)
+\vert\Delta v\vert.(1-\xi)
\nonumber\\
 &&\underset{\vert\Delta v\vert\rightarrow +\infty}{\longrightarrow}
F^{\{N,\xi\}}\Big(1\pm (1-\xi)\Big)
=(1-\xi)U.
\ee
Since, according to 
Eq.~(\ref{eq:Ts_ana_AND_eexx}),
$
T^{\{N,\xi\}}_{\rm s}(1\pm (1-\xi))=0,
$ and $E_{\rm x}^{\{N,\xi\}}(1\pm (1-\xi))=(1-\xi)U-E_{\rm H}(n)$, we
conclude that
\be\label{eq:no_corr_ener_border}
E_{\rm c}^{\{N,\xi\}}\Big(1\pm (1-\xi)\Big)=0.
\ee

\subsection{Correlation energy and potential for the $N$-centered ensemble with $\xi=1/2$}
\label{sec:Ec=0}

In the particular case $\xi_- = \xi_+ = \xi = 1/2$, the
Legendre--Fenchel transform in Eq.~(\ref{eq:LF_F_hubbard_simplified})
becomes   
\begin{eqnarray}\label{eq:LF_xi_1_over2}
F^{\{N,\xi=1/2\}}(n)
& = &  \sup_{\Delta v}
\Bigg \lbrace \varepsilon_{\rm H}(\Delta v) + \dfrac{U}{2}- \Delta v (1 - n) \Bigg \rbrace \nonumber \\
& = & T_{\rm s}^{\{N,\xi=1/2\}}(n) + \dfrac{U}{2},
\end{eqnarray}
where we used the fact that $T_{\rm
s}^{\{N,\xi=1/2\}}(n)=F^{\{N,\xi=1/2\}}(U=0,n)$. Interestingly, we first
notice that the interacting and non-interacting functionals will have
the same maximizing potential, thus leading to 
\be
\Delta v_{\rm Hxc}^{\{N,\xi=1/2\}}(n) &=&
 \Delta v_{\rm KS}^{\{N,\xi=1/2\}}(n)
-\Delta v^{\{N,\xi=1/2\}}(n) 
\nonumber\\
&=& 0.
\ee
Since, according to Eqs.~(\ref{eq:Vphys_dF_over_dn}), (\ref{eq:KS_pot})
and (\ref{eq:Ts_ana_AND_eexx}), 
\be
\Delta v_{\rm Hx}^{\{N,\xi\}}(n)&=&-\dfrac{\partial}{\partial
n}\left[E_{\rm H}(n)+E_{\rm x}^{\{N,\xi\}}(n)\right]
\nonumber\\
&=&\dfrac{U(1-2\xi)(1-n)}{(1-\xi)^2},
\ee
we conclude that 
\be\label{eq:corr_pot_zero_xi_1over2}
\Delta v_{\rm c}^{\{N,\xi=1/2\}}(n)=0.
\ee
Moreover,
since
%\begin{eqnarray}
$E_{\rm x}^{\{N,\xi=1/2\}}(n) = 
{U}/{2}-E_{\rm H}(n)$, we finally deduce from Eq.~(\ref{eq:LF_xi_1_over2}) that 
%\end{eqnarray}
\be
E_{\rm c}^{\{N,\xi=1/2\}}(n)=0.
\ee

\subsection{GACE integrand}
\label{sec:integrand_Hubbard}

According to Eqs.~(\ref{eq:GACE_integrand}), (\ref{eq:E0N-}), and
(\ref{eq:E0N+}), the GACE integrand can be
calculated as follows, 
\begin{eqnarray}\label{eq:general_gace_integrand_dimer}
&&\Delta_{\rm xc}^{\{N,\xi\}}(n)  =  U - 2E_0^N\left(\Delta
v^{\{N,\xi\}}(n)\right)
\nonumber\\
&&+ 2\varepsilon_{\rm H}\left(\Delta
v^{\{N,\xi\}}(n)\right)% \nonumber \\
+ 2\varepsilon_{\rm H}\left(\Delta v_{\rm KS}^{\{N,\xi\}}(n)\right),
%\nonumber \\
\end{eqnarray}
where $\Delta
v^{\{N,\xi\}}(n):=\Delta
v^{\{N,\underline{\xi}\}}(n)$ [we denote $\underline{\xi}=(\xi,\xi)$] is obtained numerically by Lieb
maximization (see Eq.~(\ref{eq:LF_F_hubbard_simplified})) and, according
to Eq.~(\ref{eq:KS_pot}),
\begin{eqnarray}\label{eq:KS_pot_one_weight}
\Delta v_{\rm KS}^{\{N,\xi\}}(n) = \dfrac{2t(n-1)}{\sqrt{(\xi - 1)^2 - (1 - n)^2}}.
\end{eqnarray}
We finally obtain from Eq.~(\ref{eq:E0N-}) the simplified expression
\be\label{eq:analytical_gace_integrand_dimer}
\Delta_{\rm xc}^{\{N,\xi\}}(n)  &=&  U - 2E_0^N\left(\Delta
v^{\{N,\xi\}}(n)\right)
\nonumber\\
&&- \sqrt{4t^2 +  \left[\Delta
v^{\{N,{\xi}\}}(n)\right]^2}
\nonumber\\
&&-2t\dfrac{(1-\xi)}{\sqrt{(\xi - 1)^2 - (1 - n)^2}}.
\ee
The EEXX-only contribution is obtained by differentiating the second
line of Eq.~(\ref{eq:Ts_ana_AND_eexx}) [where $\xi_+=\xi_-=\xi$] with
respect to $\xi$, thus leading to
\begin{eqnarray}\label{eq:EEXX-only_integrand}
\Delta_{\rm x}^{\{N,\xi\}}(n) & := & 
\dfrac{\partial E_{\rm x}^{\{N,\underline{\xi}\}}(n)}{\partial \xi}
= U\dfrac{\xi(n - 1)^2}{(\xi-1)^3}.
\end{eqnarray}
As expected, the latter expression gives a good approximation to the xc
GACE integrand in the weakly-correlated regime (see the top panel of
Fig.~\ref{fig:integrand}).  
Note also that, for a given density $n$ and any value of $U/t$, the correlation GACE integrand
becomes zero when approaching the border of the non-interacting $v$-representability
domain, i.e. when 
$n\rightarrow 1\pm(1-\xi)$ or, equivalently, $\xi\rightarrow 1\pm(1-n)$.
%$\xi=1\pm(1-n)$
This can be related to
Eq.~(\ref{eq:no_corr_ener_border}) which, after differentiation with
respect to $\xi$ [note that the infinitesimal variation $\xi\rightarrow\xi-\eta$
where $\eta\rightarrow0^+$ should be considered in order to
differentiate the functional within the representability domain], gives
\be
&&\Delta_{\rm c}^{\{N,\xi\}}\Big(1\pm
(1-\xi)\Big)=\left.\dfrac{\partial E_{\rm
c}^{\{N,\xi\}}(n)}{\partial\xi}\right|_{n=1\pm
(1-\xi)}
\nonumber\\
&&=\pm\left.\dfrac{\partial E_{\rm
c}^{\{N,\xi\}}(n)}{\partial n}\right|_{n=1\pm
(1-\xi)}
\nonumber\\
&&=\mp\Delta v_{\rm c}^{\{N,\xi\}}\Big(1\pm
(1-\xi)\Big)
.\ee    
According to Eq.~(\ref{eq:corr_pot_zero_xi_1over2}), the latter quantity is indeed equal to zero
when $\xi=1/2$. Numerical values of the
correlation potential obtained by Lieb maximization confirm that this
statement holds for
$\xi<1/2$, which is in complete agreement with all panels in
Fig.~\ref{fig:integrand}.  

\subsection{IP from the shifted HOMO energy and the DD}\label{sec:IP}
%%%%%%%%% to be polished ....%%%

In order to compute each contribution to the IP expression in
Eq.~(\ref{eq:exact_IP_exp_xi_zero}) separately, the $N$-centered analog
of the Levy--Zahariev shift should be calculated first. From
Eq.~(\ref{eq:densfun_Levy_shift}) and the
second-quantized expression for local potentials in the two-site model
(see Eq.~(\ref{eq:Hamil_Hubbard_dimer_model})), we obtain  
\begin{eqnarray}
C^{\{N,{\bm \xi}\}}(n) = \dfrac{1}{2} \left[E_{\rm Hxc}^{\{N,{\bm
\xi}\}}(n) -(1-n) \Delta v_{\rm Hxc}^{\{N,{\bm \xi}\}}(n)\right]. \nonumber \\
\end{eqnarray}
Turning to the DD, it comes from Eq.~(\ref{eq:LF_F_hubbard_simplified}),
\be
\dfrac{\partial F^{\{N,{\bm \xi}\}}(n)}{\partial \xi_-}=
\left[\varepsilon_{\rm
H}(\Delta v)-\dfrac{1}{2}E_0^{N}(\Delta v)\right]_{\Delta v=\Delta
v^{\{N,{\bm\xi}\}}(n)}
.
\ee
Since $\partial T_{\rm s}^{\{N,{\bm \xi}\}}(n)/\partial
\xi_-=0$, we conclude that  
\be
\dfrac{\partial E^{\{N,{\bm \xi}\}}_{\rm xc}(n)}{\partial \xi_-}=
\left[\varepsilon_{\rm
H}(\Delta v)-\dfrac{1}{2}E_0^{N}(\Delta v)\right]_{\Delta v=\Delta
v^{\{N,{\bm\xi}\}}(n)}
.
\ee
Note that, when ${\bm\xi}=0$ and $n=n_{\Psi_0^N}$ [or, equivalently, $\Delta
v^{\{N,{\bm\xi}=0\}}(n)=\Delta v_{\rm ext}$],
\be
\tilde{\varepsilon}_{\rm H}^{\{N,{\bm \xi}=0\}}=E_0^{N}(\Delta v_{\rm ext})/2,
\ee
and
\be
E_0^{N_-}(\Delta v_{\rm ext})=\varepsilon_{\rm
H}(\Delta v_{\rm ext}),
\ee
so that Eq.~(\ref{eq:exact_IP_exp_xi_zero}) is recovered, as expected.
\iffalse%%%%%%%%%%%%%
\be
\dfrac{\partial T_{\rm s}^{\{N,{\bm \xi}\}}(n)}{\partial \xi_-}=\left[\varepsilon_{\rm
H}(\Delta v)-\dfrac{1}{2}E_0^{N}(\Delta v)\right]_{\Delta v=\Delta
v^{\{N,{\bm\xi}\}}(n)}
\ee
\fi%%%%%%%%%%%% 

%%%%%%%%%%%%% END %%%%%%%%%%%%%%
\iffalse%%%%
as well as the noninteracting one [Eq.~(\ref{eq:Ts_LF})]:
\begin{eqnarray}\label{eq:LF_Ts_hubbard}
T_{\rm s}^{\{N,{\bm \xi}\}}(n) = \sup_{\Delta v} 
\Big\lbrace \mathcal{E}_{\rm KS}^{\{N,{\bm \xi}\}}(\Delta v) 
-{\Delta v}(1 - n)\Big\rbrace.
\end{eqnarray}
The noninteracting ensemble energy is given by Eq.~(\ref{eq:nonint_energy_N=2})
where
\begin{eqnarray}\label{eq:HOMO}
 \varepsilon_{\rm L}^{N}(\Delta v) = 
 -  \varepsilon_{\rm H}^{N}(\Delta v) = 
 \sqrt{t^2 + (\Delta v^2 / 4)},
\end{eqnarray}

Now that we have access to the interacting as well as noninteracting
individual energies analytically, thus also the ensemble energies,
the maximising interacting potential $\Delta v^{\{N,\xi\}}[n]$
as well as the maximising noninteracting one $\Delta v^{\{N,\xi\}}_{\rm KS}[n]$ in Eqs.~(\ref{eq:LF_F_hubbard}) and (\ref{eq:LF_Ts_hubbard}),
respectively, are easily obtained, leading to the
exact ensemble Hxc potential:
\begin{eqnarray}
\Delta v_{\rm Hxc}^{\{N,{\bm \xi}\}}(n) =\Delta v_{\rm KS}^{\{N,{\bm \xi}\}}(n) - \Delta v^{\{N,{\bm \xi}\}}(n),
\end{eqnarray}
as well as the exact ensemble Hxc energy functional
\begin{eqnarray}
E_{\rm Hxc}^{\{N,\bm{\xi}\}}(n) = F^{\{N,\bm{\xi}\}}(n) - T_{\rm s}^{\{N,\bm{\xi}\}}(n).
\end{eqnarray}
This $N$-centered ensemble Hxc energy functional
obtained by Lieb maximization is plotted in Fig.~\ref{fig:EHxc} in
the case $\xi_-=\xi_+=\xi$.
\fi%%%%%%%%%%%%%%%%%%%%%%%%%%%
\newcommand{\Aa}[0]{Aa}

\end{document}